\documentclass[sigconf, nonacm]{acmart}
\usepackage{pvldb}

\usepackage{float}
\usepackage{symbols}
\usepackage{lipsum}
\usepackage{cleveref}

\usepackage{booktabs}   
\usepackage{multirow}   
\usepackage{pifont}
\usepackage{enumitem}
\newcommand{\cmark}{\ding{51}}%
\newcommand{\xmark}{\ding{55}}%
\usepackage{amsthm}
\usepackage{xparse}
\newtheorem{problem}{Problem}
\setlist[enumerate,1]{leftmargin=0.7cm,label=\arabic*.}
\setlist[itemize]{leftmargin=0.4cm}

\newcommand{\namebase}{HorizANN}
\newcommand{\name}{\namebase\xspace}
\newcommand{\system}{\textsf{\namebase}\xspace}
\newcommand{\harmony}{\textsf{Harmony}\xspace}
\newcommand{\sptag}{\textsf{SPTAG}\xspace}
\NewDocumentCommand{\gmm}{g}{IBM\IfValueT{#1}{#1}\xspace}

\newcommand*\circled[1]{\tikz[baseline=(char.base)]{
            \node[shape=circle,fill,inner sep=1pt] (char) {
            \textcolor{white}{#1}
            } ;}}
\newcommand{\ignore}[1]{{}}
\newcommand{\cut}[1]{{}}

\definecolor{Acolor}{HTML}{944003}
\definecolor{Bcolor}{HTML}{7678ED}
\definecolor{Ccolor}{HTML}{213F80}
\definecolor{Dcolor}{HTML}{8ECAE6}
\definecolor{Ecolor}{HTML}{2A9D8F}
\definecolor{Fcolor}{HTML}{E9C46A}
\definecolor{Gcolor}{HTML}{F4A261}
\definecolor{Hcolor}{HTML}{E76F51}

\definecolor{cAmain}{HTML}{1f77b4}  
\colorlet{cAlight}{cAmain!25}
\definecolor{cBmain}{HTML}{ff7f0e}  
\colorlet{cBlight}{cBmain!25}
\definecolor{cCmain}{HTML}{2ca02c}  
\colorlet{cClight}{cCmain!25}
\definecolor{cDmain}{HTML}{d62728}  
\colorlet{cDlight}{cDmain!25}
\definecolor{cEmain}{HTML}{9467bd}  
\colorlet{cElight}{cEmain!25}
\definecolor{cFmain}{HTML}{8c564b}  
\colorlet{cFlight}{cFmain!25}
\definecolor{cGmain}{HTML}{e377c2}  
\colorlet{cGlight}{cGmain!25}
\definecolor{cZmain}{HTML}{030303}  
\colorlet{cZlight}{cZmain!25}
\colorlet{cZlightlight}{cZmain!5}
\definecolor{cPositivemain}{HTML}{2ca02c}  
\colorlet{cPositivelight}{cPositivemain!25}
\definecolor{cNegativemain}{HTML}{d62728}  
\colorlet{cNegativelight}{cNegativemain!25}

\definecolor{dillcolor}{HTML}{1f77b4}
\colorlet{dillcolorlight}{dillcolor!25}
\definecolor{shevcolor}{HTML}{2ca02c}
\colorlet{shevcolorlight}{shevcolor!25}
\definecolor{zospcolor}{HTML}{ff7f0e}
\colorlet{zospcolorlight}{zospcolor!25}
\definecolor{zodbcolor}{HTML}{e377c2}
\colorlet{zodbcolorlight}{zodbcolor!25}
\definecolor{criucolor}{HTML}{9467bd}
\colorlet{criucolorlight}{criucolor!25}
\definecolor{exhaustcolor}{HTML}{030303}
\colorlet{exhaustcolorlight}{exhaustcolor!25}
\definecolor{pgacolor}{HTML}{d62728}
\colorlet{pgacolorlight}{pgacolor!25}

\colorlet{horizanncolor}{cAmain}   
\colorlet{harmonycolor}{cBmain}    
\colorlet{sptagcolor}{cCmain}      

\pgfplotsset{
  horizann/.style={horizanncolor, thick, mark=*},
  harmony/.style={harmonycolor, thick, mark=square*},
  sptag/.style={sptagcolor, thick, mark=triangle*},
}

\colorlet{siftcolor}{cEmain}       
\colorlet{spacevcolor}{cGmain}     
\colorlet{gmmcolor}{cFmain}        

\pgfplotsset{
  dataset sift/.style={siftcolor, thick, solid,
      mark=o, mark size=1.4pt},
  dataset spacev/.style={spacevcolor, thick, dashed,
      mark=square, mark size=1.4pt, mark options={solid}},
  dataset gmm/.style={gmmcolor, thick, dotted,
      mark=triangle, mark size=1.7pt, mark options={solid}},
}

\colorlet{limitcolor}{cDmain}          
\colorlet{insertcolor}{horizanncolor}  
\colorlet{rebalcolor}{cDmain}          

\pgfplotsset{
  failure mark/.style={only marks, mark=x, mark size=3pt, very thick},
  harmony oom/.style={forget plot, failure mark, harmonycolor},
  sptag timeout/.style={forget plot, failure mark, sptagcolor},
}

\newcount\paretoK    
\newcount\paretoN    
\newcount\paretoBad  
\pgfkeys{/pareto/.cd,
    select/.store in=\paretoM,
    from/.store in=\paretofrom,
    x/.store in=\paretoxcol,
    y/.store in=\paretoycol,
    where/.store in=\paretowherecol,
    equals/.store in=\paretowhereval,
    oom/.store in=\paretooom,
}
\tikzset{
    pareto oom drop/.style={densely dashed, thin},   
    pareto oom label/.style={font=\scriptsize, inner sep=1pt,
        anchor=south east, xshift=-3.5pt, yshift=1pt},
}
\pgfplotsset{
    pareto oom mark/.style={forget plot, clip=false, failure mark},
    pareto/.style={
        x filter/.code={%
            \let\paretoorig\pgfmathresult
            \pgfmathfloatparsenumber{\paretoorig}\let\paretox\pgfmathresult
            \pgfmathfloatlessthan{\paretomax}{\paretox}%
            \ifpgfmathfloatcomparison
                \global\let\paretomax\paretox
                \pgfmathfloatlessthan{\paretox}{\paretofromF}%
                \ifpgfmathfloatcomparison
                    \let\pgfmathresult\empty      
                \else
                    \def\paretokeep{0}%
                    \ifnum\paretoM=0 \def\paretokeep{1}\fi
                    \edef\paretoloop{\noexpand\pgfplotsforeachungrouped
                        \noexpand\paretotgt\space in {\paretotargets}}%
                    \paretoloop{\ifnum\paretotgt=\paretoK\relax\def\paretokeep{1}\fi}%
                    \global\advance\paretoK by 1
                    \ifnum\paretokeep=1 \let\pgfmathresult\paretoorig
                    \else \let\pgfmathresult\empty \fi
                \fi
            \else
                \let\pgfmathresult\empty
            \fi
        },
    },
}
\newcommand{\paretoclean}[3]{%
    \pgfplotstablecreatecol[create col/assign/.code={%
        \def\paretoskip{0}%
        \ifx\paretowherecol\empty\else
            \getthisrow{\paretowherecol}\paretowherecell
            \edef\paretowhereA{\paretowherecell}%
            \edef\paretowhereB{\paretowhereval}%
            \ifx\paretowhereA\paretowhereB\else\def\paretoskip{1}\fi
        \fi
        \ifnum\paretoskip=1
            \pgfkeyssetvalue{/pgfplots/table/create col/next content}{#3}%
        \else
            \getthisrow{#1}\paretocell
            \pgfkeys{/pgf/fpu/handlers/invalid number/.code=
                {\pgfmathfloatcreate{3}{0.0}{0}}}
            \pgfmathfloatparsenumber{\paretocell}%
            \pgfmathfloatgetflagstomacro\pgfmathresult\paretoflag
            \ifnum\paretoflag=3
                \global\advance\paretoBad by 1
                \pgfkeyssetvalue{/pgfplots/table/create col/next content}{#3}%
            \else
                \pgfkeyslet{/pgfplots/table/create col/next content}\paretocell
            \fi
        \fi
    }]{#2}\paretoraw
}
\newcommand{\paretoplot}[3][]{%
    \pgfkeys{/pareto/.cd, select=0, from=0, x=recall_at_10, y=qps,
        where=, equals=, oom=, #1}%
    \pgfplotstableread[col sep=comma]{#3}\paretoraw
    \global\paretoBad=0
    \paretoclean{\paretoxcol}{paretox}{nan}%
    \paretoclean{\paretoycol}{paretoy}{0}%
    \pgfplotstablesort[sort key=paretoy, sort cmp=float >]{\paretosorted}{\paretoraw}%
    \pgfmathfloatparsenumber{\paretofrom}\global\let\paretofromF\pgfmathresult
    \global\paretoN=0
    \pgfmathfloatparsenumber{0}\global\let\paretomax\pgfmathresult
    \pgfplotstableforeachcolumnelement{paretox}\of\paretosorted\as\paretocell{%
        \pgfmathfloatparsenumber{\paretocell}\let\paretox\pgfmathresult
        \pgfmathfloatlessthan{\paretomax}{\paretox}%
        \ifpgfmathfloatcomparison
            \global\let\paretomax\paretox
            \pgfmathfloatlessthan{\paretox}{\paretofromF}%
            \ifpgfmathfloatcomparison\else
                \global\advance\paretoN by 1
                \xdef\paretolastx{\paretocell}%
                \pgfplotstablegetelem{\pgfplotstablerow}{paretoy}\of\paretosorted
                \global\let\paretolasty\pgfplotsretval
            \fi
        \fi
    }%
    \gdef\paretotargets{-1}%
    \ifnum\paretoM=1 \gdef\paretoM{2}\fi   
    \ifnum\paretoM>1
        \ifnum\paretoN>\paretoM
            \pgfmathtruncatemacro\paretolast{\paretoM-1}%
            \foreach \i in {0,...,\paretolast}{%
                \pgfmathtruncatemacro\paretot{floor(\i*(\paretoN-1)/(\paretoM-1)+0.5)}%
                \xdef\paretotargets{\paretotargets,\paretot}%
            }%
        \else \gdef\paretoM{0}\fi   
    \fi
    \global\paretoK=0
    \pgfmathfloatparsenumber{0}\global\let\paretomax\pgfmathresult
    \addplot[#2, pareto] table[x=paretox, y=paretoy] {\paretosorted};
    \ifx\paretooom\empty\else\ifnum\paretoBad>0 \ifnum\paretoN>0
        \edef\paretodo{%
            \noexpand\draw[\paretooom, pareto oom drop]
                (axis cs:\paretolastx,\paretolasty) --
                (axis cs:\paretolastx,\pgfkeysvalueof{/pgfplots/ymin});
            \noexpand\addplot[\paretooom, pareto oom mark]
                coordinates {(\paretolastx,\pgfkeysvalueof{/pgfplots/ymin})};
            \noexpand\node[pareto oom label, text=\paretooom]
                at (axis cs:\paretolastx,\pgfkeysvalueof{/pgfplots/ymin}) {OOM};
        }\paretodo
    \fi\fi\fi
}

\tikzset{
    barzero/.style={%
        draw=cZmain, fill=white,
        pattern color=cZmain, postaction={pattern=grid},
    },
    >=LaTeX
}
\tikzset{
    barone/.style={%
        draw=cAmain, fill=white,
        pattern color=cAmain, postaction={pattern=crosshatch},
    },
    >=LaTeX
}
\tikzset{
    bartwo/.style={%
        draw=cCmain, fill=white,
        pattern color=cCmain, postaction={pattern=crosshatch dots},
    },
    >=LaTeX
}
\tikzset{
    barthree/.style={%
        draw=cBmain, fill=white,
        pattern color=cBmain, postaction={pattern=north west lines},
    },
    >=LaTeX
}
\tikzset{
    barfour/.style={%
        draw=cGmain, fill=white,
        pattern color=cGmain, postaction={pattern=north east lines},
    },
    >=LaTeX
}
\tikzset{
    barfive/.style={%
        draw=cFmain, fill=white,
        pattern color=cFmain, postaction={pattern=grid},
    },
    >=LaTeX
}
\tikzset{
    barsix/.style={%
        draw=cEmain, fill=white,
        pattern color=cEmain, postaction={pattern=horizontal lines},
    },
    >=LaTeX
}
\tikzset{
    barmain/.style={%
        draw=pgacolor, fill=pgacolorlight,
        pattern color=pgacolor, postaction={pattern=crosshatch},
    },
    >=LaTeX
}

\newcommand\artifactavailabilityurl{https://github.com/illinoisdata/HorizANN-VLDB2027}

\newif\ifdraft
\drafttrue  

\begin{document}
\title{\namebase: Elastically Scaling Vector Search with Ghost Edges}

\author{Talika Gupta}
\authornote{Both authors contributed equally to this work.}
\affiliation{%
  \institution{Univ.~of Illinois Urbana-Champaign}
  \country{}
}
\email{talikag2@illinois.edu}

\author{Yunqi Li}
\authornotemark[1]
\affiliation{%
  \institution{Univ.~of Illinois Urbana-Champaign}
  \country{}
}
\email{yunqili4@illinois.edu}

\author{Aditya Kunte}
\affiliation{%
  \institution{Univ.~of Illinois Urbana-Champaign}
  \country{}
}
\email{akunte2@illinois.edu}

\author{Claire Lin}
\affiliation{%
  \institution{Univ.~of Illinois Urbana-Champaign}
  \country{}
}
\email{csl12@illinois.edu}

\author{Supawit Chockchowwat}
\affiliation{%
  \institution{CMKL University, Thailand}
  \country{}
}
\email{supawit@cmkl.ac.th}

\author{Daniel Waddington}
\affiliation{%
  \institution{IBM Research, USA}
  \country{}
}
\email{daniel.waddington@ibm.com}

\author{Hubertus Franke}
\affiliation{%
  \institution{IBM Research, USA}
  \country{}
}
\email{frankeh@us.ibm.com}

\author{Yongjoo Park}
\affiliation{%
  \institution{Univ.~of Illinois Urbana-Champaign}
  \country{}
}
\email{yongjoo@illinois.edu}

\begin{abstract}
Approximate nearest neighbor search (ANNS) is fundamental to modern data-intensive AI applications such as retrieval-augmented generation, recommendation systems, and semantic search. Among existing indexing methods, graph-based indexes emerge as the state of the art due to their superior recall-throughput tradeoffs. Scaling these indexes to multiple machines requires partitioning the data and building indexes on each. However, when new machines are added or removed, the data must be repartitioned and the indexes reconstructed from scratch, making horizontal scaling costly at billion scale. In this work, we present \system, a distributed graph-based ANNS system that scales horizontally without requiring full index reconstruction. To achieve this, we introduce ghost edges, a mechanism for preserving cross-partition connectivity as partition boundaries change during repartitioning. By reusing the structural information that would otherwise be discarded during repartitioning, \system avoids costly full index reconstruction.
As the index grows from 100M vectors on one worker to 1B vectors on ten workers, \system inserts each 100M vectors within 2.5 hours and sustains 3.4--3.5K queries per second at 0.998 Recall@10.

\end{abstract}

\maketitle

\pagestyle{plain}
\begingroup\small\noindent\raggedright\textbf{Artifact Availability:}\\
The source code, data, and/or other artifacts have been made available at \url{\artifactavailabilityurl}.
\par\endgroup

\section{Introduction}
\label{sec:intro}

\begin{table*}[t]
\centering
\caption{Comparison with prior systems. \system is the only graph-based distributed system supporting incremental indexing, online horizontal scaling and dynamic partitioning without index reconstruction.}
\label{tab:related_work}
\vspace{-2mm}
\small
\begin{tabular}{@{}p{1.2cm}lllcccp{3.2cm}@{}}
\toprule
\textbf{Group} & \textbf{System} & \textbf{Index} & \textbf{Partitioning} & \textbf{Dyn. Repart.} & \textbf{Incr. Index} & \textbf{Scale-out} & \textbf{Key Limitation} \\
\midrule
\multirow{5}{*}{\parbox{1.2cm}{\raggedright Single-Node}}
& SPANN & IVF (disk) & -- & \xmark & \xmark & -- & Static partitions \\
& SPFresh & IVF (disk) & LIRE split/merge & \cmark & \cmark & -- & Single-node only \\
& Ada-IVF & IVF & Lazy, query-aware & \cmark & \cmark & -- & Single-node only \\
& FreshDiskANN & Graph & -- & \xmark & \cmark & -- & Single-node only \\
& OdinANN & Graph & -- & \xmark & \cmark & -- & Single-node only \\
\midrule
\multirow{5}{*}{\parbox{1.2cm}{\raggedright Distributed}}
& DistributedANN & Graph (Vamana) & Random KV sharding & \xmark & \cmark & \xmark & Full KV resharding \\
& SPIRE & Hier.\ IVF & Hierarchical, LIRE & \xmark & \cmark & \xmark & Graph only for routing \\
& Harmony & Hybrid (vec + dim) & Static & \xmark & \xmark & \cmark & No graph index; no updates \\
& CoTra & Graph (proximity) & Static & \xmark & \xmark & \xmark & Static; RDMA-dependent \\
\cmidrule{2-8}
& \textbf{\system (Ours)} & \textbf{Graph (HNSW)} & \textbf{Dynamic (elastic)} & \cmark & \cmark & \cmark & -- \\
\bottomrule
\end{tabular}
\end{table*}

Modern AI applications---ranging from large-scale recommendation engines~\cite{covington2016youtube, huang2020embedding} to retrieval-augmented generation (RAG) pipelines~\cite{lewis2020rag, borgeaud2022retro,guu2020realm}---rely fundamentally on Approximate Nearest Neighbor Search (ANNS) to navigate billions of high-dimensional vectors through various types of indexes such as KD-trees~\cite{bentley1975kdtree} and random projection trees~\cite{dasgupta2008random}, hash-based methods such as locality-sensitive hashing~\cite{indyk1998approximate, andoni2015practical}, cluster-based methods such as inverted file indexes (IVF)~\cite{jegou2010product}, and graph-based methods (e.g., HNSW~\cite{hnsw}, NSG~\cite{nsg}).
To handle these immense workloads, state-of-the-art production systems such as Milvus~\cite{milvus} and Weaviate~\cite{weaviate} partition datasets into subindexes and distribute them across clusters on each worker node. When a system indexes millions of new vectors daily or faces highly variable workloads, expanding or shrinking cluster capacity without service interruption is critical.

However, these existing systems are \emph{horizontally inelastic}: they cannot dynamically add or remove worker nodes. The core problem with existing distributed graph-based systems is that \emph{they statically assign vectors to worker nodes}, partitioning data permanently at construction time. Whenever worker nodes are added or removed, the new vector-worker assignment shifts partition boundaries, invalidating the local graph structures and forcing subindexes to undergo full reconstruction from scratch~\cite{sptag, milvus}. At billion-scale, this full index rebuilding takes hours and frequently requires partitions to be taken offline~\cite{simhadri2022resultsneurips21challengebillionscale, spfresh}.\footnote{%
    A naive way to reduce downtime is to block inserts and let a replica serve queries while full repartitioning runs. However, this complicates the system design and raises operational costs.
}
Because current architectures lack a mechanism to dynamically redistribute data across an evolving topology, distributed graph indexes remain bottlenecked by a costly choice between resource underutilization and severe service downtime.

To eliminate this inelasticity, our goal is to design a distributed graph-based ANNS system that supports \emph{dynamic repartitioning}---the ability to seamlessly redistribute vector data across an altering number of worker nodes without full index reconstruction. Realizing such an elastic, horizontally scalable architecture, however, requires overcoming two fundamental challenges:
\textbf{(1) Structural Dependency of Graph-based Index.} Unlike cluster-based indexes (such as IVF)~\cite{jegou2010product, spann, spfresh, adaivf} where independent vectors are mapped strictly to independent centroids and can be predictably moved together, vectors in a graph index are bound by structural proximity edges established through expensive beam search~\cite{hnsw, dong2011efficient}. Naively repartitioning across a new topology discards these cross-boundary edges, and re-establishing them later requires repeating the costly search phase entirely.
\textbf{(2) Multi-Worker Query Routing Bottlenecks.} To maintain high throughput during cluster resizing, queries must be routed selectively. While broadcasting to all workers eliminates the performance benefits of partitioning~\cite{query_router_graph_partitioning}, under-routing or inaccurate partition selection risks missing relevant local structures and severely degrading recall~\cite{baranchuk2018revisiting}.

To address these challenges, this work introduces \system, a distributed graph-based ANNS system that scales horizontally without index reconstruction (\cref{tab:related_work}).
\textbf{(1)} \system tackles the first challenge---structural dependency---through a structure-preserving repartitioning mechanism that maintains both intra- and inter-partition edge information. When partition boundaries shift during a cluster resize, edges crossing the boundaries are preserved as \emph{ghost edges} rather than being discarded. These ghost edges are safely stored but bypassed during search, imposing zero query overhead. When vectors are subsequently redistributed, and the endpoints of a ghost edge reside in the same partition, the edge is reactivated, restoring structural graph connectivity without re-executing a beam search. 
\textbf{(2)} To address the second challenge---multi-worker query routing, \system maintains a lightweight \emph{seed index} built over a representative subset of the data. Queries first probe this seed index to accurately identify relevant partitions and are then executed exclusively on those partitions, avoiding global broadcast while preserving high recall.

In summary, we make the following contributions:
\begin{enumerate}[nosep]
    \item \textbf{Horizontally Scalable Graph Index.} We present \system, an incremental distributed graph-based ANNS system that scales horizontally. As the index grows from 100M vectors on one worker to 1B vectors on ten workers, \system sustains 3.4--3.5K QPS at Recall@10 of 0.998.
    \item \textbf{Horizontal Index Repartitioning.} Based on the concept of ghost edges, we introduce a \emph{split-and-merge} algorithm that repartitions graph indexes without reconstruction by preserving cross-partition connectivity. As the index grows to 500M vectors on 6 workers, repartitioning takes only 15\% of the total build time.
    \item \textbf{Implementation and Evaluation.} We provide an open-source implementation~\cite{horizann-code} and evaluate \system on three 1B-vector datasets in a distributed cluster, showing efficient incremental insertion with dynamic repartitioning within 2.5 hours per 100M vectors, while baselines fail to reach 1B.
\end{enumerate}

\noindent
The rest of this paper is organized as follows.
\Cref{sec:motivate} motivates our work, and \cref{sec:sys} overviews \system.
\Cref{sec:build} describes distributed indexing and querying, and \cref{sec:scale} presents split-and-merge repartitioning.
\Cref{sec:exp} evaluates \system, \cref{sec:related_work} discusses related work, and \cref{sec:conclude} concludes.

\section{Motivation}
\label{sec:motivate}

\begin{figure*}[t]
\centering
\tikzset{
    vertex/.style={circle, draw=cZmain, minimum size=5mm, inner sep=0pt, font=\scriptsize},
    vertex_a/.style={vertex, fill=cAlight, draw=cAmain},
    vertex_b/.style={vertex, fill=cBlight, draw=cBmain},
    vertex_c/.style={vertex, fill=cClight, draw=cCmain},
    unreach/.style={dash pattern=on 2pt off 1.2pt, fill=white, text=cZmain!55},
    active_edge/.style={-, semithick, draw=cZmain!85},
    cross_edge/.style={-, semithick, draw=cDmain},
    missing_edge/.style={-, draw=cDmain, line width=0.9pt, dash pattern=on 1.2pt off 1.4pt},
    arrow/.style={->, thick, >=stealth, draw=cZmain},
    label_style/.style={font=\scriptsize},
    title_style/.style={font=\small\bfseries},
    box/.style={draw=cZmain!75, line width=1pt, rounded corners},
    worker_label/.style={font=\footnotesize\bfseries, fill=white, draw=cZmain!75,
        line width=0.6pt, rounded corners=2pt, inner sep=1.5pt},
}
\newcommand{\rbworkers}[2]{%
    \node[title_style, anchor=base, text depth=0.3ex] at (2.75,-0.4) {#1};
    \foreach \xa/\xb/\name in {#2} {
        \draw[box] (\xa,0) rectangle (\xb,2.85);
        \node[worker_label] at ({(\xa+\xb)/2},2.85) {\name};
    }}
\newcommand{\rbthree}[1]{\rbworkers{#1}{0/1.7/Worker 1, 1.9/3.6/Worker 3, 3.8/5.5/Worker 2}}
\newcommand{\rbtwo}[1]{\rbworkers{#1}{0/2.68/Worker 1, 2.82/5.5/Worker 2}}

{\small
\tikz[baseline=-0.5ex]{\draw[active_edge] (0,0) -- (0.4,0);} Intra-partition \quad
\tikz[baseline=-0.5ex]{\draw[cross_edge] (0,0) -- (0.4,0);} Cross-partition \quad
\tikz[baseline=-0.5ex]{\draw[missing_edge] (0,0) -- (0.4,0);} Missing nearest-neighbor \quad
\tikz[baseline=-0.5ex]{\node[vertex, unreach, minimum size=4mm] {};} Unreachable}

\vspace{2pt}

\begin{subfigure}[t]{0.31\linewidth}
\centering
\begin{tikzpicture}[x=0.92cm, y=1cm]
\rbthree{Complete Rebuild on 3 Workers}
\node[vertex_a] (v1) at (0.39,1.22) {$v_{1}$};
\node[vertex_a] (v2) at (0.68,0.45) {$v_{2}$};
\node[vertex_a] (v3) at (1.12,2.08) {$v_{3}$};
\node[vertex_a] (v4) at (1.31,1.00) {$v_{4}$};
\node[vertex_c] (v5) at (2.31,2.22) {$v_{5}$};
\node[vertex_c] (v6) at (2.27,1.51) {$v_{6}$};
\node[vertex_c] (v7) at (3.20,1.54) {$v_{7}$};
\node[vertex_b] (v8) at (4.25,0.97) {$v_{8}$};
\node[vertex_b] (v9) at (4.85,2.18) {$v_{9}$};
\node[vertex_b] (v10) at (5.13,0.69) {$v_{10}$};
\draw[active_edge] (v1) -- (v2);
\draw[active_edge] (v1) -- (v3);
\draw[active_edge] (v1) -- (v4);
\draw[active_edge] (v2) -- (v4);
\draw[active_edge] (v3) -- (v4);
\draw[active_edge] (v5) -- (v6);
\draw[active_edge] (v5) -- (v7);
\draw[active_edge] (v6) -- (v7);
\draw[active_edge] (v8) -- (v9);
\draw[active_edge] (v8) -- (v10);
\draw[active_edge] (v9) -- (v10);
\end{tikzpicture}
\caption{Ideal}
\label{fig:repartitioning-break:ideal}
\end{subfigure}%
\hfill
\begin{subfigure}[t]{0.66\linewidth}
\centering
\begin{tikzpicture}[x=0.92cm, y=1cm]
\begin{scope}[name prefix=b-]
    \rbtwo{Before: 2 Workers}
    \node[vertex_a] (v1) at (0.39,1.22) {$v_{1}$};
    \node[vertex_a] (v2) at (0.68,0.45) {$v_{2}$};
    \node[vertex_a] (v3) at (1.12,2.08) {$v_{3}$};
    \node[vertex_a] (v4) at (1.31,1.00) {$v_{4}$};
    \node[vertex_c] (v5) at (2.31,2.22) {$v_{5}$};
    \node[vertex_c] (v6) at (2.27,1.51) {$v_{6}$};
    \node[vertex_c] (v7) at (3.20,1.54) {$v_{7}$};
    \node[vertex_b] (v8) at (4.25,0.97) {$v_{8}$};
    \node[vertex_b] (v9) at (4.85,2.18) {$v_{9}$};
    \node[vertex_b] (v10) at (5.13,0.69) {$v_{10}$};
    \draw[active_edge] (v1) -- (v2);
    \draw[active_edge] (v1) -- (v3);
    \draw[active_edge] (v1) -- (v4);
    \draw[active_edge] (v2) -- (v4);
    \draw[active_edge] (v3) -- (v4);
    \draw[active_edge] (v3) -- (v5);
    \draw[active_edge] (v4) -- (v6);
    \draw[active_edge] (v5) -- (v6);
    \draw[active_edge] (v7) -- (v8);
    \draw[active_edge] (v7) -- (v9);
    \draw[active_edge] (v8) -- (v9);
    \draw[active_edge] (v8) -- (v10);
    \draw[active_edge] (v9) -- (v10);
    \draw[cross_edge] (v5) -- (v7);
    \draw[cross_edge] (v6) -- (v7);
\end{scope}
\draw[arrow] (5.75,1.3) -- (6.35,1.3);
\node[label_style] at (6.05,1.55) {Migrate};
\begin{scope}[xshift=6.6*0.92cm, name prefix=c-]
    \rbthree{After: 3 Workers}
    \node[vertex_a] (v1) at (0.39,1.22) {$v_{1}$};
    \node[vertex_a] (v2) at (0.68,0.45) {$v_{2}$};
    \node[vertex_a] (v3) at (1.12,2.08) {$v_{3}$};
    \node[vertex_a] (v4) at (1.31,1.00) {$v_{4}$};
    \node[vertex_c] (v5) at (2.31,2.22) {$v_{5}$};
    \node[vertex_c] (v6) at (2.27,1.51) {$v_{6}$};
    \node[vertex_c, unreach] (v7) at (3.20,1.54) {$v_{7}$};
    \node[vertex_b] (v8) at (4.25,0.97) {$v_{8}$};
    \node[vertex_b] (v9) at (4.85,2.18) {$v_{9}$};
    \node[vertex_b] (v10) at (5.13,0.69) {$v_{10}$};
    \draw[active_edge] (v1) -- (v2);
    \draw[active_edge] (v1) -- (v3);
    \draw[active_edge] (v1) -- (v4);
    \draw[active_edge] (v2) -- (v4);
    \draw[active_edge] (v3) -- (v4);
    \draw[active_edge] (v5) -- (v6);
    \draw[active_edge] (v8) -- (v9);
    \draw[active_edge] (v8) -- (v10);
    \draw[active_edge] (v9) -- (v10);
    \draw[missing_edge] (v5) -- (v7);
    \draw[missing_edge] (v6) -- (v7);
\end{scope}
\end{tikzpicture}
\caption{Naive repartitioning}
\label{fig:repartitioning-break:naive}
\end{subfigure}

\vspace{-2mm}
\caption{Repartitioning a graph index from 2 to 3 workers. Vertex colors mark the worker each
vector belongs to after repartitioning.
(a)~Ideal: each worker rebuilds its graph from scratch: every vector links to its nearest
neighbors on that worker.
(b)~Naive repartitioning keeps only the edges between vectors on the same worker: it does not
maintain cross-partition edges (red), which link nearest neighbors on different workers.
When repartitioning moves both ends of a cross-partition edge onto one worker, that
nearest-neighbor edge is missing there (dotted). Here $v_7$ is left with no edges on Worker~3,
so greedy routing cannot reach it (dashed).
}
\label{fig:repartitioning-break}
\end{figure*}
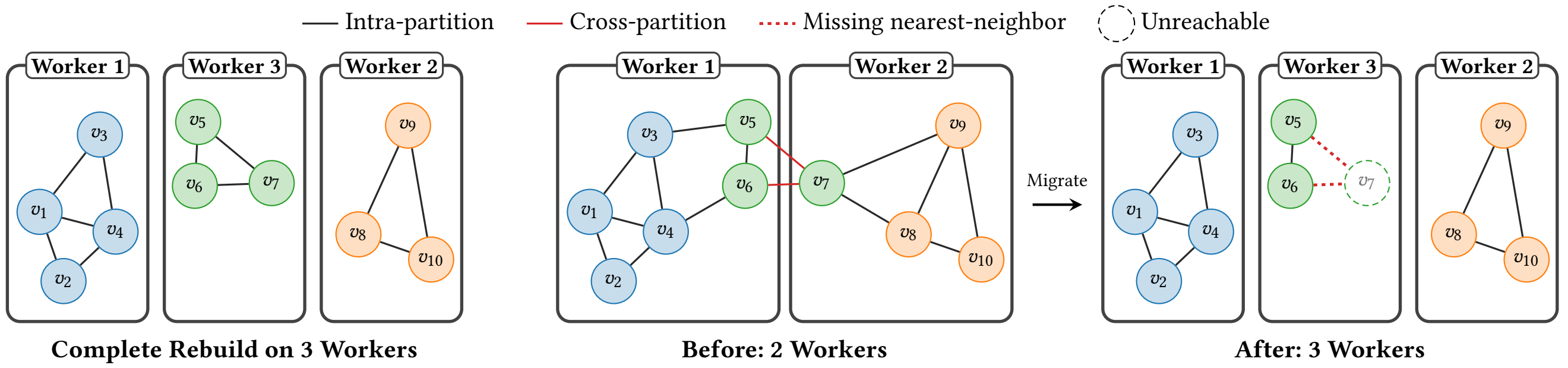


We motivate the design of \system by demonstrating the importance of graph-based indexes (\cref{sec:motivate-recall}) and highlighting the fundamental limitations of conventional index repartitioning (\cref{sec:motivate-rebalance}).

\subsection{The Necessity of Distributed Graphs}
\label{sec:motivate-recall}

Graph-based vector indexes are one of ANNS, critical for the retrieval stage of many applications like recommendation systems, semantic search, cybersecurity, and bioinformatics~\cite{graphvector}.
In particular, retrieval-augmented generation (RAG) has become the standard architecture for grounding\footnote{%
    \emph{Grounding} means that a language model gives evidence-based answers.
} large language models in external knowledge~\cite{lewis2020rag}. 
We hence quantify how downstream task quality relies on retrieval recall by evaluating 1,000 queries from HotpotQA~\cite{yang2018hotpotqa}---a multi-hop benchmark requiring two supporting documents---using Gemini 2.5 Flash. 
We simulate varying retrieval qualities by randomly introducing irrelevant documents to question answering.

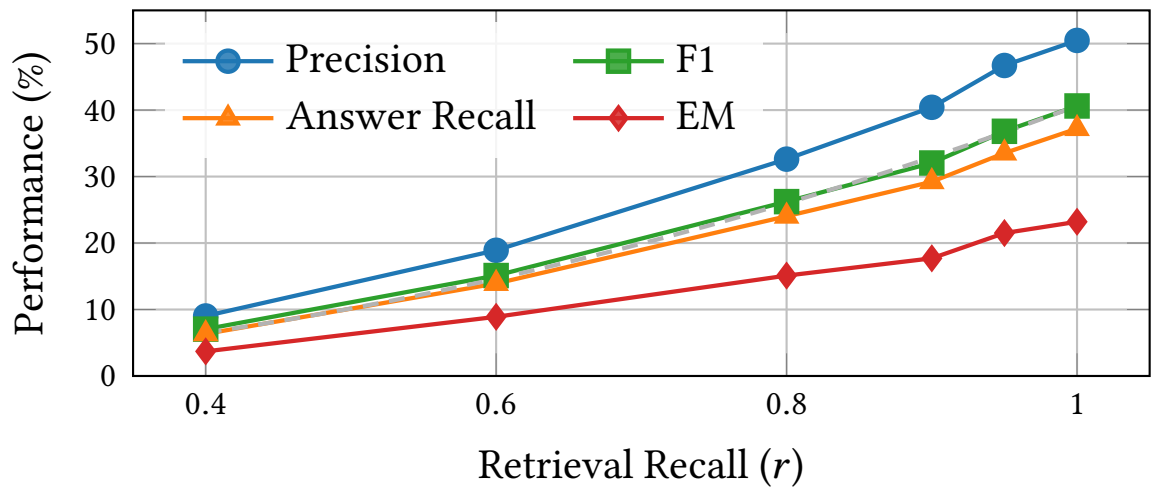
\begin{figure}[t]
\centering
\begin{tikzpicture}
\begin{axis}[
    width=0.98\linewidth,
    height=40mm,
    xlabel={Retrieval Recall ($r$)},
    ylabel={Performance (\%)},
    xmin=0.35, xmax=1.05,
    ymin=0, ymax=55,
    xtick={0.4, 0.6, 0.8, 1.0},
    ytick={0, 10, 20, 30, 40, 50},
    grid=major,
    grid style={cZlight},
    tick label style={font=\scriptsize},
    label style={font=\small},
    legend style={font=\small, at={(0.03,0.97)}, anchor=north west, draw=none, fill=white, fill opacity=0.85, text opacity=1, /tikz/every even column/.append style={column sep=0.5em}},
    legend columns=2,
    legend cell align=left,
    cycle list name=color list,
]
\addplot[cAmain, thick, mark=*] table[col sep=comma, x=recall, y=precision] {figures/1_intro/recall_vs_qa_plot.csv};
\addlegendentry{Precision}

\addplot[cCmain, thick, mark=square*] table[col sep=comma, x=recall, y=f1] {figures/1_intro/recall_vs_qa_plot.csv};
\addlegendentry{F1}

\addplot[cBmain, thick, mark=triangle*] table[col sep=comma, x=recall, y=ans_recall] {figures/1_intro/recall_vs_qa_plot.csv};
\addlegendentry{Answer Recall}

\addplot[cDmain, thick, mark=diamond*] table[col sep=comma, x=recall, y=em] {figures/1_intro/recall_vs_qa_plot.csv};
\addlegendentry{EM}

\addplot[gray!60, thick, densely dashed, domain=0.4:1.0, samples=20, forget plot] {40.62 * x^2};
\end{axis}
\end{tikzpicture}
\vspace{-2mm}
\caption{Retrieval recall vs.\ downstream QA performance on HotpotQA (1,000 queries). Curves exhibit quadratic decay ($\approx C \cdot r^2$, dashed grey curve shown for F1), confirming that multi-hop downstream reasoning collapses rapidly under sub-optimal retrieval recall.}
\label{fig:recall_vs_qa_plot}
\end{figure}

As shown in \Cref{fig:recall_vs_qa_plot}, downstream QA accuracy (EM, i.e., exact match) scales \emph{quadratically} ($r^2$) with retrieval recall $r$ due to the multi-hop dependency. 
At $r = 0.90$, EM drops from a baseline of 23.2\% down to 17.7\% (approximating $0.90^2 = 81\%$ of maximum effectiveness). More generally, tasks requiring $k$ independent pieces of evidence exhibit an $r^k$ performance decay, making ultra-high recall non-negotiable as application reasoning complexity grows.
Therefore, the strict recall-accuracy relationship calls for the adoption of graph-based indexes like HNSW, which can offer higher accuracy than other methods (e.g., IVF)~\cite{aumuller2020annbenchmarks}, when all graph data can fit in memory.


Albeit the necessity, dynamically scaling graph-based indexes into multiple worker nodes remains under-explored.
In distributed environments, horizontally scaling cluster-based indexes such as IVF variants is well-understood~\cite{spfresh, adaivf} because vectors are mapped independently to disconnected centroids.
On the other hand, handling inter-partition edges for graph-based indexes is non-trivial: vectors are explicitly bound by proximity edges. 
Filling the architectural gap between graph-level retrieval performance and cluster-level horizontal elasticity remains an open and critical problem.

\subsection{The Breakdown of Naive Repartitioning}
\label{sec:motivate-rebalance}

To illustrate why cluster resizing breaks graph structures, consider an initial two-worker distributed HNSW cluster where a third worker is introduced to balance the storage (\cref{fig:repartitioning-break}). For clarity, the example shows a single graph layer; every layer of HNSW breaks in the same way. To avoid a full global index reconstruction, an elastic system must partition local graphs and migrate the designated subgraphs to the new worker. This process directly fractures the index structure.

Under a centralized setting, a graph-based index operates by navigating a network of \emph{proximity edges}. Queries traverse from an entry point through the graph, moving progressively closer to the target vector by following these edges to nearer neighbors. While the specific algorithms used to construct these edges vary across different graph indexes, they all share this fundamental reliance on continuous edge connectivity for successful routing. However, when graphs are repartitioned across a distributed cluster, this core routing mechanism is directly disrupted.

\paragraph{Problem 1: Fractured Routing and Dropped Recall.} As conceptualized in \Cref{fig:repartitioning-break}, proximity edges frequently cross the boundaries of the newly assigned partitions. When data points migrate, these cross-boundary edges are severed. For instance, if vectors move to a new worker node, their existing incoming and outgoing edges to vectors remaining on the original worker are discarded. Because graph-based ANNS's greedy routing relies entirely on continuous edge connectivity to navigate closer to a query's neighborhood, these severed boundaries form routing dead ends. When a query searches these fractured, disjoint subgraphs, it cannot traverse the missing links, causing retrieval recall to collapse toward zero.

\paragraph{Problem 2: Catastrophic Reconstruction Costs.} Re-establishing the missing graph connectivity through standard means requires re-inserting the migrated vectors, incurring unnecessary recomputation costs. This requires re-running expensive beam searches for every vector in the altered partitions to locate valid proximity neighbors. At billion-scale, this index reconstruction takes hours which may degrade overall system availability or double the resource allocation.

An elastic ANNS system must therefore find a way to preserve cross-partition proximity edges across dynamic cluster shifts without executing recurrent, full-scale graph builds. \system achieves this balance by natively retaining these cross-partition boundaries during repartitioning.

\section{System Overview}
\label{sec:sys}

\begin{figure}[t]
    \centering
    \begin{tikzpicture}[
        scale=0.92,
        every node/.style={transform shape},
        component/.style={draw=cZmain!80, rounded corners=2.5pt, minimum height=0.55cm, 
                          minimum width=2.1cm, align=center, font=\scriptsize,
                          fill=white},
        worker_box/.style={draw=cZmain!70, rounded corners=4pt, minimum height=1.75cm, 
                           minimum width=2.1cm, align=center, fill=white},
        hnsw_box/.style={draw=cBmain, rounded corners=2.5pt, minimum height=0.7cm, 
                         minimum width=1.75cm, align=center, fill=cBlight!30,
                         font=\scriptsize\bfseries, text=cBmain!90!black},
        client/.style={draw=cZmain!80, rounded corners=3pt, minimum height=1.4cm, 
                       minimum width=1.1cm, align=center, font=\small\bfseries,
                       fill=white},
        main_box/.style={draw=cZmain!70, rounded corners=4pt, fill=white},
        search/.style={-{Stealth[scale=0.8]}, thick, cAmain},
        searchline/.style={thick, cAmain},
        insert/.style={-{Stealth[scale=0.8]}, thick, cCmain, densely dashed},
        insertline/.style={thick, cCmain, densely dashed},
        stepnum/.style={circle, fill=cAmain, text=white, font=\scriptsize\bfseries, 
                        inner sep=0.5pt, minimum size=10pt},
        stepnum_i/.style={circle, fill=cCmain, text=white, font=\scriptsize\bfseries, 
                          inner sep=0.5pt, minimum size=10pt},
        plane_ctrl/.style={draw=cAmain!40, dashed, rounded corners=5pt, fill=cAlight!15, line width=0.7pt},
        plane_data/.style={draw=cBmain!40, dashed, rounded corners=5pt, fill=cBlight!15, line width=0.7pt}
    ]

    \node[client] (client) at (0, 0) {Client};

    \node[main_box, minimum width=2.4cm, minimum height=2.6cm] (main) at (2.9, 0) {};
    \node[font=\footnotesize\bfseries, text=cAmain] at (2.9, 1.05) {Main Node};
    \node[component] (router) at (2.9, 0.5) {Query Router};
    \node[component] (agg) at (2.9, -0.15) {Aggregator};
    \node[component, minimum height=0.45cm, font=\scriptsize, minimum width=1.9cm] (meta) at (2.9, -0.8) {Metadata Mgr};

    \node[worker_box] (w1) at (6.3, 0.95) {};
    \node[font=\scriptsize\bfseries, text=cBmain] at (6.3, 1.62) {Worker ($w_1$)};
    \node[hnsw_box] (hnsw1) at (6.3, 1.1) {HNSW Index};
    \node[component, minimum width=1.6cm, minimum height=0.35cm, font=\scriptsize] (split1) at (6.3, 0.42) {Repartitioner};

    \node at (6.3, -0.05) {$\vdots$};

    \node[worker_box] (w2) at (6.3, -1.15) {};
    \node[font=\scriptsize\bfseries, text=cBmain] at (6.3, -0.48) {Worker ($w_n$)};
    \node[hnsw_box] (hnsw2) at (6.3, -1.0) {HNSW Index};
    \node[component, minimum width=1.6cm, minimum height=0.35cm, font=\scriptsize] (split2) at (6.3, -1.68) {Repartitioner};

    \begin{scope}[on background layer]
        \node[plane_ctrl, fit=(main), inner sep=7pt, inner ysep=11pt] (control_boundary) {};
        \node[font=\scriptsize\bfseries, text=cAmain!80] at ([yshift=6pt]control_boundary.south) {Control Plane};
        
        \node[plane_data, fit=(w1)(w2)(split2), inner sep=6pt, inner ysep=11pt] (data_boundary) {};
        \node[font=\scriptsize\bfseries, text=cBmain!80] at ([yshift=6pt]data_boundary.south) {Data Plane};
    \end{scope}

    \draw[searchline] (client.east |- 0, 0.5) -- (router.west);
    \draw[search] (1.45, 0.5) -- (router.west);
    \node[stepnum] at (1.1, 0.68) {1};

    \draw[searchline] (router.east) -- (4.65, 0.5);
    \draw[search] (4.65, 0.5) |- (hnsw1.west);
    \draw[search] (4.65, 0.5) |- (hnsw2.west);
    \node[stepnum] at (4.65, 0.72) {2};

    \draw[searchline] (hnsw1.west |- 6.3, 0.85) -- (4.95, 0.85);
    \draw[searchline] (hnsw2.west |- 6.3, -1.25) -- (4.95, -1.25);
    \draw[searchline] (4.95, 0.85) -- (4.95, -0.15);
    \draw[searchline] (4.95, -1.25) -- (4.95, -0.15);
    \draw[search] (4.95, -0.15) -- (agg.east);
    \node[stepnum] at (4.4, -0.15) {3};

    \draw[search] (agg.west) -- (client.east |- 0, -0.15);
    \node[stepnum] at (1.1, -0.15) {4};

    \draw[insertline] (client.north) -- (0, 2.05) -- (2.7, 2.05);
    \draw[insert] (2.7, 2.05) -- (2.7, 1.3);
    \node[stepnum_i] at (1.35, 2.05) {1};

    \draw[insertline] (3.1, 1.3) -- (3.1, 2.35) -- (6.3, 2.35);
    \draw[insert] (6.3, 2.35) -- (w1.north);
    \node[stepnum_i] at (4.7, 2.35) {2};

    \draw[insertline] (6.6, 1.83) -- (6.6, 2.65) -- (-0.85, 2.65) -- (-0.85, 0.2);
    \draw[insert] (-0.85, 0.2) -- (client.west |- 0, 0.2);
    \node[stepnum_i] at (2.9, 2.65) {3};

    \node[font=\scriptsize, anchor=north] at (2.3, -2.1) {
        \tikz[baseline=-0.5ex]{\draw[search] (0,0) -- (0.5,0);} \textsc{Search}$(q,K)$
        \quad\quad
        \tikz[baseline=-0.5ex]{\draw[insert] (0,0) -- (0.5,0);} \textsc{Insert}$(v)$
    };

    \end{tikzpicture}
    \caption{\system system architecture. The \textbf{Control Plane} (Main Node) coordinates request routing, aggregation, and metadata. The \textbf{Data Plane} (Workers) maintains local HNSW partitions and repartitioning logic. Solid and dashed arrows denote query and insertion data flows, respectively.}
    \label{fig:architecture_diagram}
\end{figure}
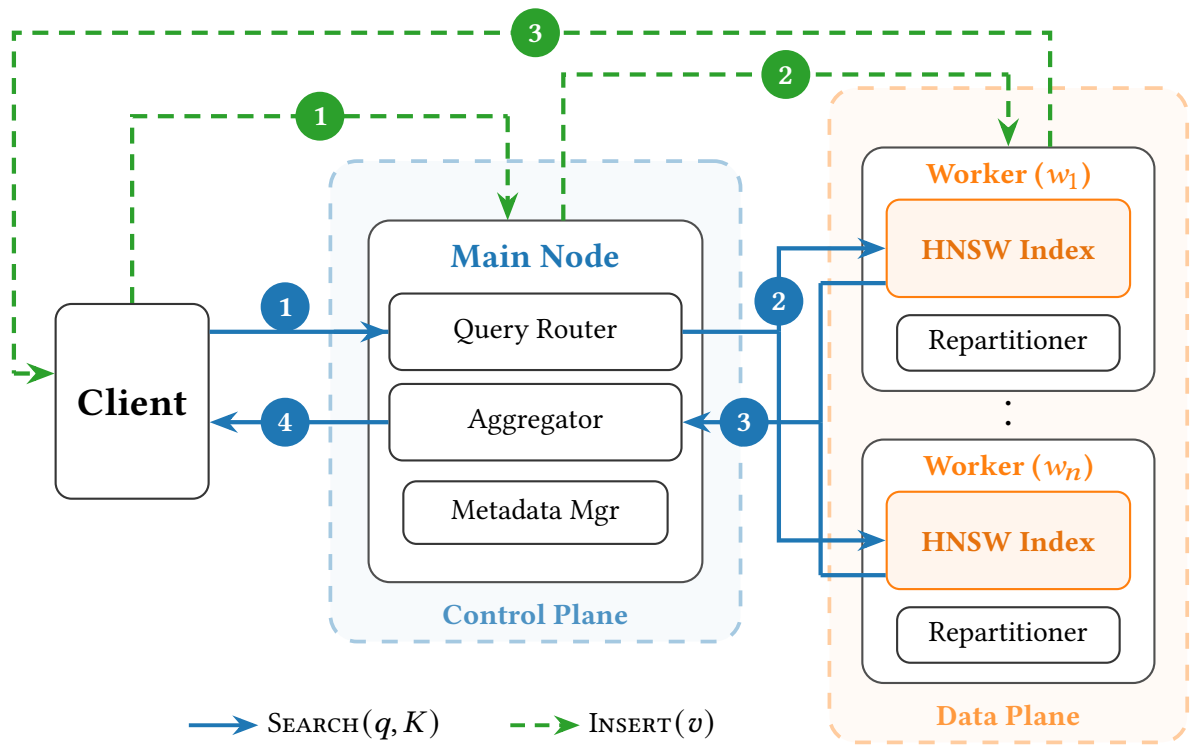

This section outlines the design of \system, defining the external user interface (\cref{sec:sys-ui}) and the internal architecture (\cref{sec:sys-arch}). We then discuss the logical data model that \system represents (\cref{sec:sys-hnsw}) and the design goals it achieves (\cref{sec:sys-properties}).

\subsection{User Interface}
\label{sec:sys-ui}

\system provides a high-level abstraction that consists of five operations.
They allow users to interact with a horizontally scalable vector index as if it were a single, local data structure. 
From the user's perspective, the system maintains a global collection of vectors $\mathcal{V} \subset \bbR^d$; we denote a vector by $v \in \bbR^d$ and a query by $q \in \bbR^d$.
This design effectively hides the complexities of the distributed architecture like query routing, data partitioning, and result aggregation.

\paragraph{Data Maintenance} Data maintenance is performed through incremental operations that keep the index synchronized with the application's evolving state. A user invokes $\textsc{Insert}(v)$ to add a new vector $v \in \bbR^d$ to the global collection $\mathcal{V}$. This is typically called when new searchable entities, such as user profiles, document embeddings, or product features, are generated. Conversely, $\textsc{Delete}(v)$ is used to remove a specific vector from $\mathcal{V}$. This ensures the vector is immediately pruned from the search space and will no longer appear in later searches.

\paragraph{Search} The primary mechanism for retrieval is the $\textsc{Search}(q,K)$ operation, which enables similarity-based discovery. By providing a query vector $q \in \bbR^d$ and a parameter $K$, the user requests the $K$ most similar vectors in $\mathcal{V}$ relative to $q$. The system identifies the global top-$K$ approximate nearest neighbors and returns them as a single ranked list.

\paragraph{Resource Management} \system also exposes administrative operations for elastic scaling. A user can call $\textsc{AddWorker}(w)$ to integrate a new compute and storage resource into the cluster. This is typically prompted by the growth of $\mathcal{V}$ toward memory limits or a need to increase search throughput under high query loads. Similarly, $\textsc{RemoveWorker}(w)$ allows for the decommissioning of a resource for cost optimization during periods of lower demand.


\subsection{System Architecture}
\label{sec:sys-arch}

\system follows a manager-worker architecture (\autoref{fig:architecture_diagram}) comprising a single \textbf{main node} that coordinates multiple \textbf{workers} to transparently expose a unified vector index.

The \textbf{main node} coordinates client requests and cluster membership across three core components:
(1)~\emph{Metadata Manager}: Maintains cluster topology, partition centers $\{c_i\}_{i=1}^k$ for the assignment function $\pi(v) = \arg\min_i \|v - c_i\|$, per-worker vector counts, and a reservoir sample of vectors for re-clustering upon scaling (\cref{sec:scale-actuator}).
(2)~\emph{Query Router}: Determines candidate partitions likely to contain the nearest neighbors of a query vector, backed by an auxiliary seed HNSW index over sampled representative points (\cref{sec:index-query}).
(3)~\emph{Aggregator}: Gathers local search candidates from probed workers and merges them into a global top-$K$ ranked result.

Each \textbf{worker} $w_i$ stores a disjoint partition $V_i$ of the vector collection $\mathcal{V}$ ($\bigcup_i V_i = \mathcal{V}$, $V_i \cap V_j = \emptyset$ for $i \neq j$) centered at $c_i$. Each worker maintains a local HNSW index $H_i$ over $V_i$ (\cref{sec:sys-hnsw}), structuring its graph into \emph{active edges} connecting co-located vectors and \emph{ghost edges} capturing cross-partition proximity relationships (\cref{sec:index-structure}). Additionally, each worker hosts a \emph{repartitioner} that executes local split and merge operations during scaling (\cref{sec:scale-mechanism}), reactivating ghost edges into active edges without full graph reconstruction.

\paragraph{Data Flow}
Request execution across these components proceeds in three flows:
(1)~\emph{Query}: The client submits $\textsc{Search}(q, K)$ to the main node. The query router queries its seed index to identify the target subset of partitions $\mathcal{P}(q)$ nearest to $q$. Probed workers execute local HNSW search in parallel, and the aggregator merges the returned local candidates into the global top-$K$ nearest neighbors returned to the client.
(2)~\emph{Insertion and Deletion}: For $\textsc{Insert}(v)$, the main node evaluates $\pi(v)$ and routes $v$ to worker $w_{\pi(v)}$ for local insertion, additionally querying neighboring workers to establish ghost edges that preserve the KNN property (\cref{sec:index-insert-delete}). For $\textsc{Delete}(v)$, the request is routed to $w_{\pi(v)}$ for lazy pruning.
(3)~\emph{Elastic Scaling}: When triggered by capacity threshold $\tau$ or administrative operations ($\textsc{AddWorker}$ / $\textsc{RemoveWorker}$), the metadata manager recomputes partition centers via balanced $k$-means. Workers execute local splits, transfer migrating child indexes, and merge incoming indexes via ghost edge reactivation (\cref{sec:scale}).

\subsection{Logical Data Model}
\label{sec:sys-hnsw}

\system logically represents a Hierarchical Navigable Small World (HNSW) graph~\cite{hnsw}, albeit a physically distributed one across workers. Note that \system's techniques also apply to other graph-based indexes such as NSG~\cite{nsg} and Vamana~\cite{diskann}.

\paragraph{Data Structure}
HNSW is a graph based index that combines small-world graphs~\cite{smallworld} and skip-lists~\cite{skiplist} to create a multi-layer structure for efficient similarity search. The HNSW index $H$ over vectors $\mathcal{V}$ consists of layers $L_0, L_1, \ldots$, with each layer being a graph $G = (V, E)$ whose vertices are vectors, i.e., $V \subseteq \mathcal{V}$.
Each edge $(u, v) \in E$ connects vectors that are likely near each other. Each $v$ is assigned a level $\ell(v)$ at insertion and appears in layers $L_0, \ldots, L_{\ell(v)}$, so $V = \mathcal{V}$ on $L_0$ and $V$ shrinks on upper layers. A vertex keeps at most $2M$ out-edges in $L_0$ and $M$ in each upper layer, and search starts from the \emph{entry point}, a vector with the highest level. Unless stated otherwise, the techniques in this paper treat every layer in the same way, so we describe them on a single layer.

\paragraph{Construction}
HNSW is built incrementally. Each inserted vector $v$ is assigned a random maximum layer $\ell(v)$. The algorithm searches for nearest neighbors at each layer from $L_{\ell(v)}$ down to $L_0$, and adds edges connecting the new node to up to $M$ neighbors per layer.

\paragraph{Search}
Given a query vector $q$, search traverses layers top-to-bottom. At each layer, greedy traversal moves to the neighbor closest to $q$ until no improvement is found; this node becomes the entry point for the next layer. Upper layers locate the neighborhood, and the bottom layer returns the final top-$K$ candidates.

\paragraph{KNN Property}
Each vertex in HNSW is connected to a subset of its approximate nearest neighbors found by beam search; we refer to this as the \emph{KNN property}. Greedy search relies on this property to move progressively closer to the query.

\subsection{Design Goals}
\label{sec:sys-properties}

\system is designed to achieve four design goals.
(i)~\textit{Consistent Accuracy}: \system as a distributed system maintains nearly consistent recall as index and cluster size grow. 
(ii)~\textit{Efficient Repartitioning}: \system quickly adapts itself with changing workers and workload, avoiding a full index reconstruction.
(iii)~\textit{Throughput Scale-up}: \system's query throughput scales with the number of workers, enabling the system to handle increased load by adding machines.
(iv)~\textit{Online Repartitioning}: \system continues to serve queries during repartitioning, sustaining throughput comparable to that before repartitioning.



\section{Horizontal Indexing}
\label{sec:build}


This section describes how \system organizes and operates its distributed index over a given set of workers. We first define the distributed index structure, whose ghost edges preserve the KNN property (\cref{sec:sys-hnsw}) across partition boundaries (\cref{sec:index-structure}). We then describe how vectors are partitioned, inserted, and deleted while preserving this property (\cref{sec:index-insert-delete}), and how a query is routed to only the relevant partitions and aggregated into a global result (\cref{sec:index-query}). How the index adapts when workers join or leave the cluster is deferred to \cref{sec:scale}.

\subsection{Distributed Index Structure}
\label{sec:index-structure}

\system partitions vectors across workers, and each worker $w_i$ maintains a local HNSW index over its partition $V_i$. Following \cref{sec:sys-hnsw}, we describe a single layer, on which $w_i$ stores a graph $G_i = (V_i, E_i)$ consisting of the out-edges of its vertices. \system divides these edges into two types:
\begin{align*}
E_i^{\text{active}} &= \{(u,v) \in E_i : v \in V_i\}, \\
E_i^{\text{ghost}} &= \{(u,v) \in E_i : v \notin V_i\}.
\end{align*}
\emph{Active edges} connect vectors on the same worker and are traversed by search. \emph{Ghost edges} connect a vector to vectors on other workers; they are stored but never traversed, so each worker searches its local HNSW index alone. Ghost edges arise in two ways: insertion connects a new vector to its nearest neighbors on other workers (\cref{sec:index-insert-delete}), and repartitioning turns an active edge into a ghost edge when its endpoints are moved apart (\cref{sec:scale}).

\subsection{Distributed Insertion and Deletion}
\label{sec:index-insert-delete}

\system supports incremental insertions and deletions of vectors to the distributed index. The distributed index consists of local HNSW indexes, one per worker, where each worker owns a spatially coherent partition of the vector space. 



\paragraph{Partitioning.}
Vectors are partitioned across workers using balanced $k$-means clustering on a sample of points sampled using reservoir sampling~\cite{reservoir_sampling}, where $k$ is the number of workers. Each worker $w_i$ is associated with a cluster center $c_i$ obtained from $k$-means. We define the \emph{partition assignment function} $\pi: \bbR^d \rightarrow \{1, \ldots, k\}$ as:
\[
\pi(v) = \arg\min_i \|v - c_i\|
\]
A vector $v$ is assigned to worker $w_{\pi(v)}$, the worker whose center is nearest to $v$. This ensures semantic locality within partitions, enabling efficient query routing (\cref{sec:index-query}).

\paragraph{Insertion.}
To insert vector $v$, the main node routes $v$ to worker $w_{\pi(v)}$, which inserts $v$ into its local HNSW index and creates active edges to $v$'s approximate nearest neighbors within the partition. A vector near a partition boundary, however, may have some of its nearest neighbors on other workers. The main node therefore also sends $v$ to the other workers selected by the query router (\cref{sec:index-query}); each of them searches its local index for $v$'s nearest neighbors, and $w_{\pi(v)}$ stores the results as ghost edges of $v$. Active and ghost edges together approximate $v$'s neighborhood across partitions, so insertion preserves the KNN property (\cref{sec:sys-hnsw}).


\paragraph{Deletion.}
To delete vector $v$, the Main node routes the request to worker $w_{\pi(v)}$. The local HNSW index lazily deletes $v$. Subsequent searches skip deleted vectors.

\paragraph{Bulk Build.}
Given a collection of $n$ vectors and $k$ workers, \system builds the distributed index by first partitioning vectors using the assignment function $\pi$, then constructing local HNSW indexes in parallel on each worker. Since partitions are disjoint, no coordination is required. Construction time is $O((n/k) \log (n/k))$ per worker. The Main node batches insertions to increase throughput, and each local HNSW index processes a received batch concurrently.

\subsection{Distributed Query}
\label{sec:index-query}

With the index partitioned across several partitions, effective querying requires \emph{finding the smallest number of partitions which contains the true nearest neighbors for a given query.} 

\begin{figure}[t]
\centering
\begin{tikzpicture}
\begin{axis}[
    width=\linewidth,
    height=48mm,
    ybar=0pt,
    bar width=6pt,
    xlabel={\# partitions containing ground-truth top-10 neighbors},
    ylabel={Queries (\%)},
    xtick={1,2,3,4,5,6},
    xmin=0.4, xmax=6.6,
    ymin=0, ymax=70,
    ytick={0,20,40,60},
    ymajorgrids,
    grid style={cZlight},
    tick label style={font=\footnotesize},
    label style={font=\small},
    legend style={font=\footnotesize, at={(0.98,0.95)}, anchor=north east, draw=none, fill=white, fill opacity=0.8, text opacity=1},
    legend cell align=left,
]
\addplot[barthree] table[col sep=comma, x=num_partitions, y expr=\thisrow{p5}/100] {figures/2_method/neighbor_distribution.csv};
\addlegendentry{5 partitions (mean 1.96)}
\addplot[barone] table[col sep=comma, x=num_partitions, y expr=\thisrow{p10}/100] {figures/2_method/neighbor_distribution.csv};
\addlegendentry{10 partitions (mean 2.73)}
\end{axis}
\end{tikzpicture}
\vspace{-2mm}
\caption{Distribution of partitions containing ground-truth top-10 neighbors on SIFT1M. Ground-truth neighbors for most queries lie in only a few partitions, with the count varying across queries. 
}
\label{fig:neighbor_distribution}
\end{figure}
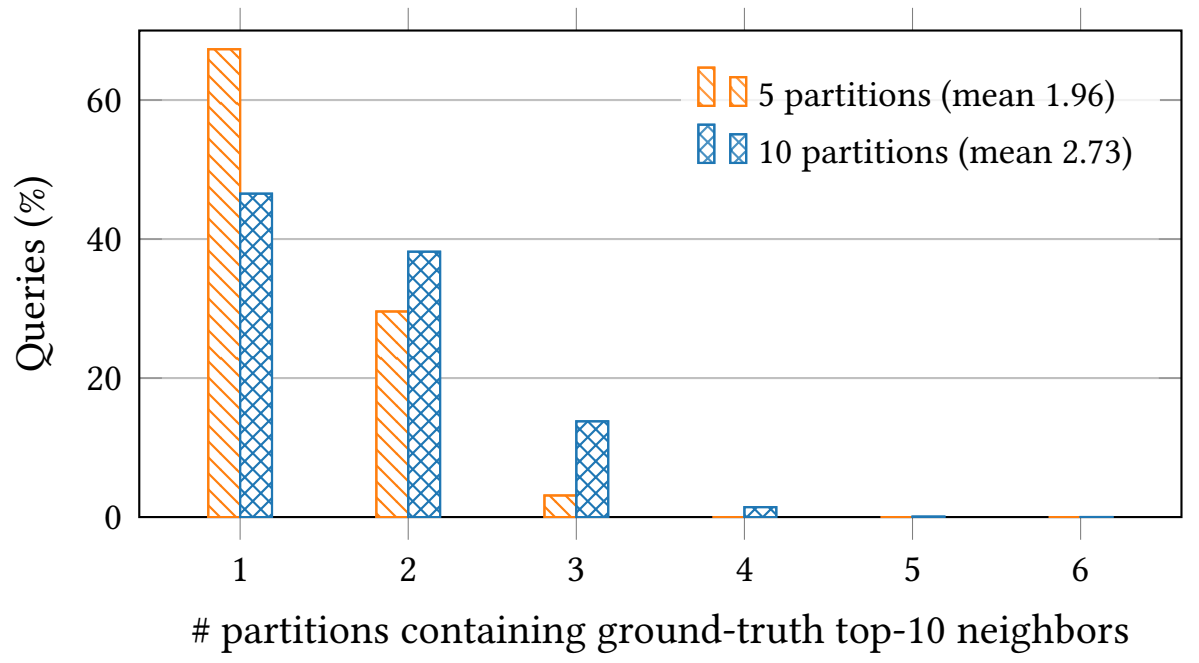

In a small scale experiment, we partitioned the SIFT1M dataset ($1$ million embeddings, $128$ dimensional) into $10$ and $5$ partitions and measured, for each query, how many partitions contain at least one true top-$10$ neighbor. \autoref{fig:neighbor_distribution} shows the results. On average, true neighbors reside in only a few partitions, indicating that a small number of probes $n_{\text{probe}}$ suffices for high recall. A constant $n_{\text{probe}}$ is also wasteful, since it varies across queries. This motivates the need for a query-aware router which routes a query $q$ to only the most relevant partitions, to avoid under or over routing. 


\paragraph{Query Router} A key challenge in distributed ANNS is routing queries to relevant partitions. For systems with many small partitions (e.g. IVF with $10^6$ partitions of $10^3$ 
points each), representing each partition by its centroid suffices. However, in our setting of fewer partitions with large sizes, centroid based routing performs poorly, as large partitions  exhibit heterogeneous internal structure that a single 
point cannot capture~\cite{query_router_graph_partitioning}.

Gottesbüren et al.~\cite{query_router_graph_partitioning} addresses this via hierarchical clustering within each partition to build representative points and indexes them for fast retrieval. We adopt a simpler approach suited to incremental settings: maintain a reservoir sample~\cite{reservoir_sampling} of data points indexed via HNSW. This seed index is the query router.

As vectors arrive, we maintain a reservoir sample $\mathcal{R} \subset \mathcal{V}$ of size $m$ (typically $0.01\%$ of the dataset). Each sampled point $v$ is annotated with its current partition assignment $\pi(v)$. This assignment changes after each repartitioning operation. We build an HNSW index $I_{\text{seed}}$ over $\mathcal{R}$ for fast approximate search.

Given a query $q \in \bbR^d$ and desired result size $K$, we first determine which partitions to probe. Let $k_{\text{seed}}$ be the retrieval parameter (typically $k_{\text{seed}} = K$). We search $I_{\text{seed}}$ for the $k_{\text{seed}}$ approximate nearest neighbors $Q \subseteq \mathcal{R}$ of $q$, then route $q$ to all partitions $\mathcal{P}(q) = \bigcup_{x \in Q} \{\pi(x)\}$. 

Our approach differs from \cite{query_router_graph_partitioning} in two aspects: (1) we use sampled data points as representatives, avoiding the cost of hierarchical k-means clustering in each partition, and (2)~repartitioning does not require recomputing representatives; reservoir sampling maintains the sample in $O(1)$ amortized time per insertion, so representatives update naturally as data arrives or partitions change. In contrast, k-means-based representatives must be recomputed from scratch.

\paragraph{Aggregator} Each probed worker performs local HNSW search and returns its local top-$K$ candidates; the \emph{Aggregator} merges these results to produce the global top-$K$ nearest neighbors.

\section{Horizontal Index Scaling}
\label{sec:scale}

While \cref{sec:build} assumes a given set of workers, \system must also adapt as workers join or leave the cluster. This capability serves two of our design goals (\cref{sec:sys-properties}): \emph{Throughput Scale-up}, since newly added workers absorb a share of the data and query load, and \emph{Efficient Repartitioning}, since the cluster must reach its new configuration without a full index reconstruction. Both hinge on \emph{repartitioning}, the process of redistributing data across a potentially changing number of workers. This section formalizes the repartitioning problem on a single layer (\cref{sec:scale-objective}), presents \system's split-and-merge mechanism, which extends naive repartitioning with ghost edges, first on a single layer and then on all layers of HNSW (\cref{sec:scale-mechanism}), describes when repartitioning is triggered and how data moves when workers are added or removed (\cref{sec:scale-actuator}), and explains how \system keeps serving queries while repartitioning is in progress (\cref{sec:scale-liveness}). 




\subsection{Problem Formulation}
\label{sec:scale-objective}

Following \cref{sec:sys-hnsw}, each local index $H_i$ consists of one graph per layer, and we write $G_i^{(l)}$ for its layer-$l$ graph. We first consider a single layer and drop the superscript, so worker $i$ holds a graph $G_i = (V_i, E_i)$ whose edges $E_i$ encode proximity relationships discovered by beam search during insertion; \cref{sec:scale-mechanism} extends repartitioning to all layers. The distributed index is then $\mathcal{G} = \{G_1, \ldots, G_k\}$ over the vector set $\mathcal{V} = \bigcup_{i=1}^{k} V_i$.
When a worker is added or removed, \system recomputes the partition centers for the new worker count $k'$ (\cref{sec:scale-actuator}), which yields a new partition assignment function $\pi'$. We formalize repartitioning as follows.

\begin{problem}[Repartitioning]\label{prob:repartitioning}
Given a distributed index $\mathcal{G} = \{G_1, \ldots, G_k\}$ over $\mathcal{V}$ and a new partition assignment function $\pi' : \bbR^d \to \{1, \ldots, k'\}$, construct a distributed index $\mathcal{G}' = \{G'_1, \ldots, G'_{k'}\}$ in which each $G'_j = (V'_j, E'_j)$ is a graph index over the vectors $V'_j = \{v \in \mathcal{V} : \pi'(v) = j\}$.
\end{problem}

A good solution is fast to compute and preserves search accuracy, that is, $\mathcal{G}'$ should be about as accurate as indexes built from scratch on $V'_1, \ldots, V'_{k'}$.
Naive repartitioning (\cref{sec:motivate-rebalance}) cannot achieve both. It drops cross-partition edges, which hurts accuracy. And restoring them requires re-inserting every affected vector, which is slow.






\subsection{Elastic Repartition Mechanism}
\label{sec:scale-mechanism}

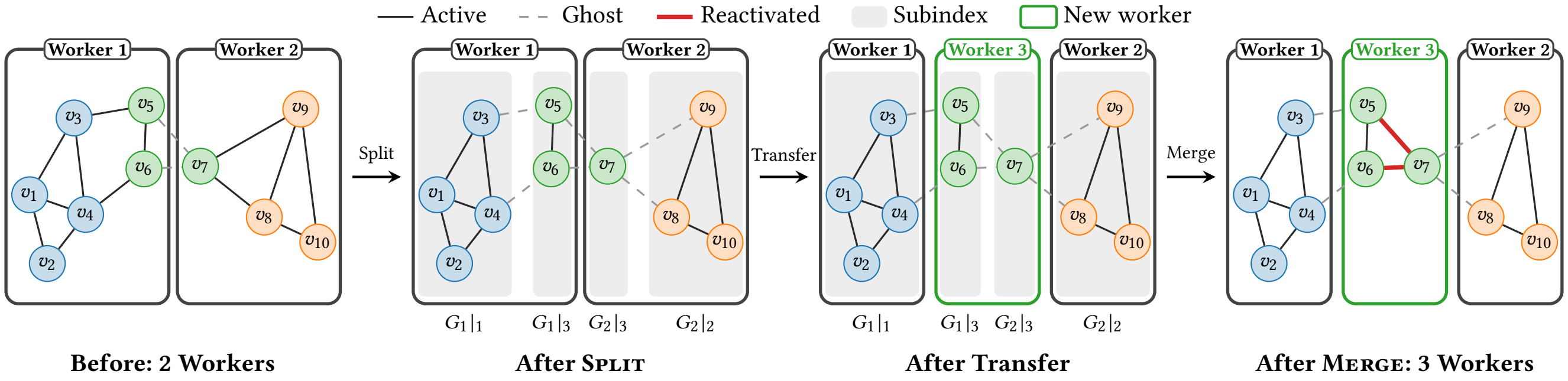
\begin{figure*}[t]
\centering
\tikzset{
    vertex/.style={circle, draw=cZmain, minimum size=4mm, inner sep=0pt, font=\scriptsize},
    vertex_a/.style={vertex, fill=cAlight, draw=cAmain},
    vertex_b/.style={vertex, fill=cBlight, draw=cBmain},
    vertex_c/.style={vertex, fill=cClight, draw=cCmain},
    active_edge/.style={-, semithick, draw=cZmain!85},
    ghost_edge/.style={-, dashed, semithick, draw=cZmain!40},
    reactivated_edge/.style={-, draw=cDmain, line width=1.4pt},
    arrow/.style={->, thick, >=stealth, draw=cZmain},
    label_style/.style={font=\scriptsize},
    title_style/.style={font=\small\bfseries},
    box/.style={draw=cZmain!75, line width=1pt, rounded corners},
    box_new/.style={draw=cCmain, line width=1pt, rounded corners},  
    worker_label/.style={font=\scriptsize\bfseries, fill=white, draw=cZmain!75,
        line width=0.6pt, rounded corners=2pt, inner sep=1pt},
    subindex/.style={fill=cZmain!7, rounded corners=2pt},
}
\newcommand{\smworkers}[2]{%
    \node[title_style, anchor=base, text depth=0.3ex] (title) at (2.75,-0.75) {#1};
    \foreach \xa/\xb/\name/\isnew in {#2} {
        \ifnum\isnew=1
            \draw[box_new] (\xa,0) rectangle (\xb,2.85);
            \node[worker_label, draw=cCmain, text=cCmain] at ({(\xa+\xb)/2},2.85) {\name};
        \else
            \draw[box] (\xa,0) rectangle (\xb,2.85);
            \node[worker_label] at ({(\xa+\xb)/2},2.85) {\name};
        \fi
    }}
\newcommand{\smtwo}[1]{\smworkers{#1}{0/2.68/Worker 1/0, 2.82/5.5/Worker 2/0}}
\newcommand{\smthree}[1]{\smworkers{#1}{0/1.7/Worker 1/0, 1.9/3.6/Worker 3/1, 3.8/5.5/Worker 2/0}}
\newcommand{\smsubindexes}[1]{%
    \foreach \xa/\xb/\name [count=\k] in {#1} {
        \fill[subindex] (\xa,0.08) rectangle (\xb,2.6);
        \node[label_style, below] (s\k) at ({(\xa+\xb)/2},0) {$\name$};
    }}
\newcommand{\smvectors}{%
    \node[vertex_a] (v1) at (0.39,1.22) {$v_{1}$};
    \node[vertex_a] (v2) at (0.68,0.45) {$v_{2}$};
    \node[vertex_a] (v3) at (1.12,2.08) {$v_{3}$};
    \node[vertex_a] (v4) at (1.31,1.00) {$v_{4}$};
    \node[vertex_c] (v5) at (2.31,2.22) {$v_{5}$};
    \node[vertex_c] (v6) at (2.27,1.51) {$v_{6}$};
    \node[vertex_c] (v7) at (3.20,1.54) {$v_{7}$};
    \node[vertex_b] (v8) at (4.25,0.97) {$v_{8}$};
    \node[vertex_b] (v9) at (4.85,2.18) {$v_{9}$};
    \node[vertex_b] (v10) at (5.13,0.69) {$v_{10}$};
    \draw[active_edge] (v1) -- (v2);
    \draw[active_edge] (v1) -- (v3);
    \draw[active_edge] (v1) -- (v4);
    \draw[active_edge] (v2) -- (v4);
    \draw[active_edge] (v3) -- (v4);
    \draw[active_edge] (v5) -- (v6);
    \draw[active_edge] (v8) -- (v9);
    \draw[active_edge] (v8) -- (v10);
    \draw[active_edge] (v9) -- (v10);}

{\small
\tikz[baseline=-0.5ex]{\draw[active_edge] (0,0) -- (0.4,0);} Active \quad
\tikz[baseline=-0.5ex]{\draw[ghost_edge] (0,0) -- (0.4,0);} Ghost \quad
\tikz[baseline=-0.5ex]{\draw[reactivated_edge] (0,0) -- (0.4,0);} Reactivated \quad
\tikz[baseline=-0.5ex]{\fill[subindex] (0,-0.12) rectangle (0.4,0.12);} Subindex \quad
\tikz[baseline=-0.5ex]{\draw[box_new, line width=0.8pt, rounded corners=1pt] (0,-0.12) rectangle (0.4,0.12);} New worker}

\vspace{2pt}
\begin{tikzpicture}[x=0.68cm, y=1cm]
\begin{scope}[name prefix=a-]
    \smtwo{Before: 2 Workers}
    \smvectors
    \draw[active_edge] (v3) -- (v5);
    \draw[active_edge] (v4) -- (v6);
    \draw[active_edge] (v7) -- (v8);
    \draw[active_edge] (v7) -- (v9);
    \draw[ghost_edge] (v5) -- (v7);
    \draw[ghost_edge] (v6) -- (v7);
\end{scope}
\draw[arrow] (5.7,1.42) -- (6.5,1.42);
\node[label_style] at (6.1,1.67) {Split};
\begin{scope}[xshift=6.7*0.68cm, name prefix=b-]
    \smtwo{After \textsc{Split}}
    \smsubindexes{0.08/1.62/G_1|_1, 1.97/2.60/G_1|_3, 2.90/3.53/G_2|_3, 3.88/5.42/G_2|_2}
    \smvectors
    \draw[ghost_edge] (v3) -- (v5);
    \draw[ghost_edge] (v4) -- (v6);
    \draw[ghost_edge] (v7) -- (v8);
    \draw[ghost_edge] (v7) -- (v9);
    \draw[ghost_edge] (v5) -- (v7);
    \draw[ghost_edge] (v6) -- (v7);
\end{scope}
\draw[arrow] (12.4,1.42) -- (13.2,1.42);
\node[label_style] at (12.8,1.67) {Transfer};
\begin{scope}[xshift=13.4*0.68cm, name prefix=c-]
    \smthree{After Transfer}
    \smsubindexes{0.08/1.62/G_1|_1, 1.97/2.64/G_1|_3, 2.86/3.53/G_2|_3, 3.88/5.42/G_2|_2}
    \smvectors
    \draw[ghost_edge] (v3) -- (v5);
    \draw[ghost_edge] (v4) -- (v6);
    \draw[ghost_edge] (v7) -- (v8);
    \draw[ghost_edge] (v7) -- (v9);
    \draw[ghost_edge] (v5) -- (v7);
    \draw[ghost_edge] (v6) -- (v7);
\end{scope}
\draw[arrow] (19.1,1.42) -- (19.9,1.42);
\node[label_style] at (19.5,1.67) {Merge};
\begin{scope}[xshift=20.1*0.68cm, name prefix=d-]
    \smthree{After \textsc{Merge}: 3 Workers}
    \smvectors
    \draw[ghost_edge] (v3) -- (v5);
    \draw[ghost_edge] (v4) -- (v6);
    \draw[ghost_edge] (v7) -- (v8);
    \draw[ghost_edge] (v7) -- (v9);
    \draw[reactivated_edge] (v5) -- (v7);
    \draw[reactivated_edge] (v6) -- (v7);
\end{scope}
\end{tikzpicture}

\vspace{-2mm}
\caption{Repartitioning with \textsc{Split} and \textsc{Merge} (\cref{alg:hnsw-repartition}) on the
example of \Cref{fig:repartitioning-break}. The left panel shows the index before
repartitioning; vertex colors show the new assignment
$\pi'$ (\cref{sec:scale-actuator}), with Worker~3 (green) as the added worker.
\textsc{Split} divides the index of each worker $W_i$ into one subindex $G_i|_j$ per destination
worker $W_j$ (shaded), and edges between different subindexes become ghost edges.
Each worker $W_i$ then transfers every subindex $G_i|_j$ with $j \neq i$ to $W_j$.
\textsc{Merge} combines the subindexes each worker receives and reactivates (red) the ghost edges
whose ends are both on it ($v_5$--$v_7$ and $v_6$--$v_7$).}
\label{fig:split_merge}
\end{figure*}

Instead of discarding cross-partition edges, \system takes a different approach: it preserves and reuses them to construct the new edge sets $\{E'_1, E'_2, \ldots, E'_{k'}\}$ at a repartitioning. 

\paragraph{Repartitioning on Single Layer}

\begin{algorithm}[t]
\caption{\system Repartitioning on Single Layer}
\label{alg:hnsw-repartition}
\SetArgSty{textnormal}
\SetFuncSty{textsc}
\SetKwFor{ParForEach}{foreach}{do in parallel}{end}
\SetKwProg{Fn}{Function}{}{}
\SetKwFunction{FnSplit}{Split}
\SetKwFunction{FnMerge}{Merge}
\KwInput{Local indexes $G_1, \ldots, G_k$ of one layer; new partition assignment function $\pi'$ (\cref{sec:scale-actuator})}
\KwOutput{New local indexes $G'_1, \ldots, G'_{k'}$}
\ParForEach{worker $i \in \{1, \ldots, k\}$}{
    $\{G_i|_1, \ldots, G_i|_{k'}\} \gets$ \FnSplit{$G_i, \pi'$}\;\label{alg:hnsw-repartition:split_call}
    Transfer each $G_i|_j$ to worker $j$\Comment*[r]{kept if $j = i$}\label{alg:hnsw-repartition:transfer_call}
}
\ParForEach{worker $j \in \{1, \ldots, k'\}$}{
    $G'_j \gets$ \FnMerge{$G_1|_j, \ldots, G_k|_j$}\;\label{alg:hnsw-repartition:merge_call}
}
\Return{$G'_1, \ldots, G'_{k'}$}\;
\BlankLine
\Fn{\FnSplit{$G_i, \pi'$}}{
    \ForEach{$j \in \{1, \ldots, k'\}$}{
        $V_i|_j \gets \{v \in V_i : \pi'(v) = j\}$\;
        $E_i|_j \gets \{(u, v) \in E_i : u \in V_i|_j\}$\;
        mark $(u, v) \in E_i|_j$ active if $v \in V_i|_j$, and ghost otherwise\;\label{alg:hnsw-repartition:split}
    }
    \Return{$G_i|_j = (V_i|_j, E_i|_j)$ for $j = 1, \ldots, k'$}\;
}
\BlankLine
\Fn{\FnMerge{$G_1|_j, \ldots, G_k|_j$}}{
    $V'_j \gets \bigcup_{i=1}^{k} V_i|_j$;\quad $E'_j \gets \bigcup_{i=1}^{k} E_i|_j$\;
    mark $(u, v) \in E'_j$ active if $v \in V'_j$, and ghost otherwise\;\label{alg:hnsw-repartition:reactivate}
    prune active edges of each vertex to the degree bound\;\label{alg:hnsw-repartition:prune}
    \Return{$G'_j = (V'_j, E'_j)$}\;
}
\end{algorithm}

\Cref{alg:hnsw-repartition} repartitions one layer $G_i$ in two steps. 
Each worker $i$ calls \textsc{Split} to divide its local index $G_i$ by destination worker (line~\ref{alg:hnsw-repartition:split_call}). Each subindex carries every outgoing edge $(u, v)$ of its vectors. $(u, v)$ stays active only if $v$ lies in the same subindex. (line~\ref{alg:hnsw-repartition:split}).
Then each worker transfers each subindex $G_i|_j$ to its destination $j$ (line~\ref{alg:hnsw-repartition:transfer_call}).
\textsc{Merge}  combines the subindexes that worker $j$ receives into $G'_j$ (line~\ref{alg:hnsw-repartition:merge_call}). It reactivates every ghost edge whose ends are now both on worker $j$ (line~\ref{alg:hnsw-repartition:reactivate}). This reconnects subindexes from different source workers without re-inserting their vectors, while the remaining ghost edges are kept for later repartitionings. 
Finally, vertices whose active degree exceeds the bound are pruned with the HNSW neighbor selection heuristic (line~\ref{alg:hnsw-repartition:prune}).
\cref{fig:split_merge} illustrates this procedure.

\paragraph{Extending to HNSW}
\system repartitions each HNSW index $H_i$ by running \cref{alg:hnsw-repartition} on every layer independently, with $G_i$ taken to be its layer-$l$ graph $G_i^{(l)}$. We write $H_i|_j$ for the subindex of $H_i$ assigned to worker $j$, which consists of the layer graphs $G_i^{(l)}|_j$.
All layers of a vector $v$ go to the same worker $\pi'(v)$, so $v$'s level $\ell(v)$ remains unchanged.
\textsc{Merge} prunes each layer to its own degree bound: $2M$ edges on $L_0$ and $M$ on the upper layers. Besides running \textsc{Split} and \textsc{Merge} on every layer, \system needs two additional steps. First, a vector in $H_i|_j$ has no active $L_0$ edge if none of its $L_0$ neighbors falls into $H_i|_j$. After the transfer, worker $j$ holds $H_i|_j$ and reinserts such vectors into it before \textsc{Merge}.
Second, each worker $j$ sets the entry point of $H'_j$ to one of its highest-level vectors.

\paragraph{Ghost Edge Storage.}
Ghost edges are stored on disk and used only during repartitioning, to save memory during normal operation. For each vertex $v$, we store pairs $(u, d(v, u))$ where $d(v,u)$ is the distance between $u$ and $v$. On each layer, a vertex keeps at most as many ghost edges as its degree bound ($2M$ on $L_0$ and $M$ above).

\subsection{Dynamic Repartition Actuators}
\label{sec:scale-actuator}

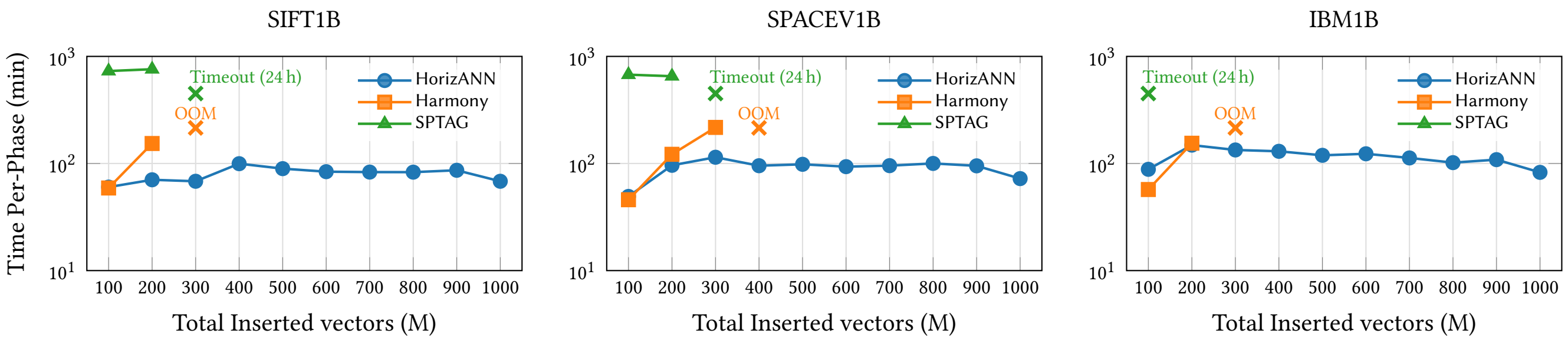
\begin{figure*}[t]
\centering
\pgfplotstableread[col sep=comma, empty cells with={nan}]
    {figures/exp/index_build/harmony_sift1b.csv}\harmonysift
\pgfplotstableread[col sep=comma, empty cells with={nan}]
    {figures/exp/index_build/harmony_spacev1b.csv}\harmonyspacev
\pgfplotstableread[col sep=comma, empty cells with={nan}]
    {figures/exp/index_build/harmony_gmm1b.csv}\harmonygmm
\newcommand{\oomy}{215}     
\newcommand{\timeouty}{450} 
\def\filternan{nan}
\newif\ifoomlabeled
\pgfplotsset{
  keep if empty/.style={
    x filter/.code={%
      \getthisrow{#1}\tmpcell
      \ifx\tmpcell\empty\else\ifx\tmpcell\filternan\else
        \def\pgfmathresult{inf}%
      \fi\fi
    },
  },
  oom label/.style={
    nodes near coords={\ifoomlabeled\else\global\oomlabeledtrue OOM\fi},
    every node near coord/.append style={failure label, text=harmonycolor,
        anchor=south, yshift=1pt},
    execute at begin plot visualization={\global\oomlabeledfalse},
  },
}
\tikzset{
  failure label/.style={font=\scriptsize, inner sep=1pt},
  timeout label/.style={failure label, text=sptagcolor,
      anchor=south west, xshift=-3pt, yshift=1pt},
}
\begin{tikzpicture}
\pgfmathsetlengthmacro{\ylabw}{1.15cm}
\pgfmathsetlengthmacro{\rpad}{0.1cm}
\pgfmathsetlengthmacro{\hsep}{0.95cm}
\pgfmathsetlengthmacro{\axw}{(\linewidth-\ylabw-\rpad-2*\hsep)/3}
\begin{groupplot}[
    group style={group size=3 by 1, horizontal sep=\hsep,
        xlabels at=edge bottom, ylabels at=edge left},
    scale only axis,
    width=\axw,
    height=2.4cm,
    xlabel={Total Inserted vectors (M)},
    ylabel={Time Per-Phase (min)},
    xmin=50, xmax=1050,
    xtick={100,200,...,1000},
    /pgf/number format/1000 sep={},
    ymode=log, ymin=10, ymax=1000,
    ytick={10,100,1000},
    log basis y=10,
    grid=major,
    grid style={gray!25},
    tick label style={font=\scriptsize},
    label style={font=\small, align=center},
    title style={font=\small},
    clip mode=individual,          
    legend pos=north east,
    legend cell align=left,
    legend style={draw=none, fill=white, fill opacity=0.8, text opacity=1,
        font=\scriptsize, nodes={inner sep=0.8pt}},
]
\nextgroupplot[title={SIFT1B}]
\addplot[horizann] table[
    col sep=comma,
    x expr=\thisrow{total_inserted}/1000000,
    y expr=\thisrow{phase_total_sec}/60,
] {figures/exp/index_build/horizann_sift1b.csv};
\addlegendentry{\system}
\addplot[harmony] table[
    x expr=\thisrow{max_base}/1000000,
    y expr=\thisrow{build_sec}/60,
] {\harmonysift};
\addlegendentry{\textsf{Harmony}}
\addplot[harmony oom, forget plot, oom label, keep if empty=build_sec] table[
    col sep=comma,
    x expr=\thisrow{max_base}/1000000,
    y expr=\oomy,
] {figures/exp/index_build/harmony_sift1b.csv};
\addplot[sptag] table[
    col sep=comma,
    x expr=\thisrow{cumulative_vectors}/1000000,
    y expr=\thisrow{phase_total_sec}/60,
] {figures/exp/index_build/sptag_sift1b.csv};
\addlegendentry{\textsf{SPTAG}}
\addplot[sptag timeout, forget plot] coordinates {(300, \timeouty)};
\node[timeout label] at (axis cs:300,\timeouty) {Timeout (24\,h)};

\nextgroupplot[title={SPACEV1B}]
\addplot[horizann] table[
    col sep=comma,
    x expr=\thisrow{total_inserted}/1000000,
    y expr=\thisrow{phase_total_sec}/60,
] {figures/exp/index_build/horizann_spacev1b.csv};
\addlegendentry{\system}
\addplot[harmony] table[
    x expr=\thisrow{max_base}/1000000,
    y expr=\thisrow{build_sec}/60,
] {\harmonyspacev};
\addlegendentry{\textsf{Harmony}}
\addplot[harmony oom, forget plot, oom label, keep if empty=build_sec] table[
    col sep=comma,
    x expr=\thisrow{max_base}/1000000,
    y expr=\oomy,
] {figures/exp/index_build/harmony_spacev1b.csv};
\addplot[sptag] table[
    col sep=comma,
    x expr=\thisrow{cumulative_vectors}/1000000,
    y expr=\thisrow{phase_total_sec}/60,
] {figures/exp/index_build/sptag_spacev1b.csv};
\addlegendentry{\textsf{SPTAG}}
\addplot[sptag timeout, forget plot] coordinates {(300, \timeouty)};
\node[timeout label] at (axis cs:300,\timeouty) {Timeout (24\,h)};

\nextgroupplot[title={\gmm{1B}}]
\addplot[horizann] table[
    col sep=comma,
    x expr=\thisrow{total_inserted}/1000000,
    y expr=\thisrow{phase_total_sec}/60,
] {figures/exp/index_build/horizann_gmm1b.csv};
\addlegendentry{\system}
\addplot[harmony] table[
    x expr=\thisrow{max_base}/1000000,
    y expr=\thisrow{build_sec}/60,
] {\harmonygmm};
\addlegendentry{\textsf{Harmony}}
\addplot[harmony oom, forget plot, oom label, keep if empty=build_sec] table[
    col sep=comma,
    x expr=\thisrow{max_base}/1000000,
    y expr=\oomy,
] {figures/exp/index_build/harmony_gmm1b.csv};
\addplot[sptag timeout, forget plot] coordinates {(100, \timeouty)};
\node[timeout label] at (axis cs:100,\timeouty) {Timeout (24\,h)};
\addlegendimage{sptag}\addlegendentry{\textsf{SPTAG}}
\end{groupplot}
\end{tikzpicture}
\vspace{-2mm}
\caption{Incremental insertion on three 1B datasets. \system maintains consistent per-phase insertion time (within 49--148 minutes). In contrast, \harmony encounters out-of-memory (OOM) errors, while \sptag experiences timeouts.}
\label{fig:eval:incremental_build_time}
\end{figure*}

\system repartitions the index in two cases: the cluster automatically adds a worker when a partition exceeds the size threshold $\tau$, and an operator can also add or remove a worker manually.


\paragraph{Worker Addition.}
Adding a worker proceeds in three steps. 
Firstly, the main node maintains a sample of vectors via reservoir sampling \cite{reservoir_sampling} from the input stream and runs balanced $k$-means on it to compute $k{+}1$ centers and broadcasts to all workers. 
Secondly, each worker derives the new assignment $\pi'$ from these centers and repartitions its index $H_i$ under $\pi'$ (\cref{sec:scale-mechanism}); only vectors whose assignment changes leave their worker.
Lastly, workers update vector counts. The main node updates the routing index to include the new worker.

\paragraph{Worker Removal.}
Removing a worker follows the similar procedure as adding with reversed data flow. The main node computes $k{-}1$ centers. The departing worker splits its index and distributes subindexes to remaining workers. After remaining workers merges received subindexes, the departing worker deregisters.

\subsection{Online Repartition}
\label{sec:scale-liveness}

To keep serving queries during repartitioning, we extend the design in \cref{sec:scale-actuator} into an online protocol, under which each query is answered entirely by the indexes either before or after repartitioning. Taking worker addition as an example, we describe the protocol below.

\paragraph{Repartitioning Protocol Design}
The key idea is to assign each query an \emph{epoch}, the version of the indexes that serves it. The main node tags each query on arrival with the epoch of its current routing index, and each worker answers the query with its index of that epoch.
While the cluster moves from epoch $e$ to $e{+}1$, the main node and every worker keep their indexes of both epochs, so queries tagged with $e$ keep using the old indexes while the new ones are built.
The protocol performs this move in three stages, starting once the workers receive the new centers (\cref{sec:scale-actuator}):

\begin{itemize}[nosep]
\item \textbf{Prepare.}
Each worker loads its ghost edges and runs \textsc{Split}, which only reads its current index $H_i$, so $H_i$ keeps serving epoch $e$; the worker writes the outgoing subindexes to disk and starts its \emph{staged index} $H'_i$ from $H_i|_i$. Meanwhile, the main node prepares the routing index of epoch $e{+}1$ by relabeling each sampled vector $x$ with $\pi'(x)$, leaving routing index of epoch $e$ unchanged.

\item \textbf{Transfer and commit.}
Once every worker has prepared, each worker $i$ transfers its outgoing subindexes and merges the incoming ones into its staged index $H'_i$ (\cref{alg:hnsw-repartition}).
Once all workers have finished merging, each worker commits its staged index $H'_i$ as its new index for epoch $e{+}1$, while keeping $H_i$ for the queries still routed under epoch $e$. 
The main node then switches the routing index to epoch $e{+}1$ and waits for the in-flight queries of $e$ to finish, after which the third stage begins.

\item \textbf{Retire and persist.}
Each worker free its old index $H_i$ and persist the ghost edges of new index $H'_i$, and the main node drops the routing index of epoch $e$ and refreshes the vector counts.
\end{itemize}

Since queries move to epoch $e{+}1$ only when the main node switches its routing index, a failure before that point is simple to undo. The main node aborts epoch $e{+}1$ and the workers drop their staged indexes $H'_i$.
 
\paragraph{Insertion during Repartitioning}
The main node buffers insertions that arrive during repartitioning and replays them through the normal insertion path (\cref{sec:index-insert-delete}) once the new indexes are committed, so that they are inserted into the new indexes and sampled into the routing index like any other insertion.

\section{Evaluation}
\label{sec:exp}

We evaluate \system on billion-scale datasets, comparing against state-of-the-art distributed vector search systems. Our experiments reveal the following findings:

\begin{table}[t]
\caption{Datasets used in evaluation.}
\label{tab:datasets}
\vspace{-2mm}
\small
\centering
\begin{tabular*}{\columnwidth}{@{\extracolsep{\fill}}llrrlr@{}}
\toprule
\textbf{Dataset} & \textbf{Source} & \textbf{Size} & \textbf{Dim.} & \textbf{Type} & \textbf{Queries} \\
\midrule
SIFT & INRIA & 1B & 128 & uint8 & 10{,}000 \\
SPACEV & Big-ANN'21 & 1B & 100 & int8 & 10{,}000 \\
\gmm{1B} & Real embeddings & 1B & 128 & int8 & 10{,}000 \\
\bottomrule
\end{tabular*}
\end{table}

\begin{itemize}[nosep]
\item \textbf{Incremental Insertion.}
During incremental index construction from 0 to 1B vectors, \system consistently inserts each successive 100M vectors in 49--148 minutes while baselines fail to insert all vectors. (\cref{sec:eval:build}) 

\item \textbf{Higher Search Performance.} 
\system achieves the highest QPS at every Recall@10 up to 0.99 on all three datasets, with 2.1--3.9$\times$ the QPS of \sptag and 9--21$\times$ that of \harmony at Recall@10 = 0.95. (\cref{sec:eval:search})

\item \textbf{Horizontal Scalability.}
As the cluster grows from 1 to 10 workers and the index grows from 100M to 1B vectors, \system achieves a 2.4--3$\times$ increase in query throughput while maintaining consistent recall across cluster sizes. (\cref{sec:eval:scalability})

\item \textbf{Repartitioning Efficiency.}
Growing the index from 0 to 500M vectors and from 1 to 6 workers, repartitioning takes 15\% of total build time, and at most 14 minutes in one pass. (\cref{sec:eval:rebalancing_efficiency})

\item \textbf{Online Repartitioning.}
\system keeps serving queries during repartitioning: when 500M vectors are repartitioned from 5 to 6 workers, throughput averages 84\% of its pre-repartitioning level and Recall@10 stays at 0.98. (\cref{sec:eval:rebalancing_liveness})

\item \textbf{Bounded Per-Worker Memory.}
As the index grows to 1B vectors, each worker's peak memory stays below 214\,GiB, and ghost edges add at most 16\% to the index memory. (\cref{sec:eval:memory})
\end{itemize}

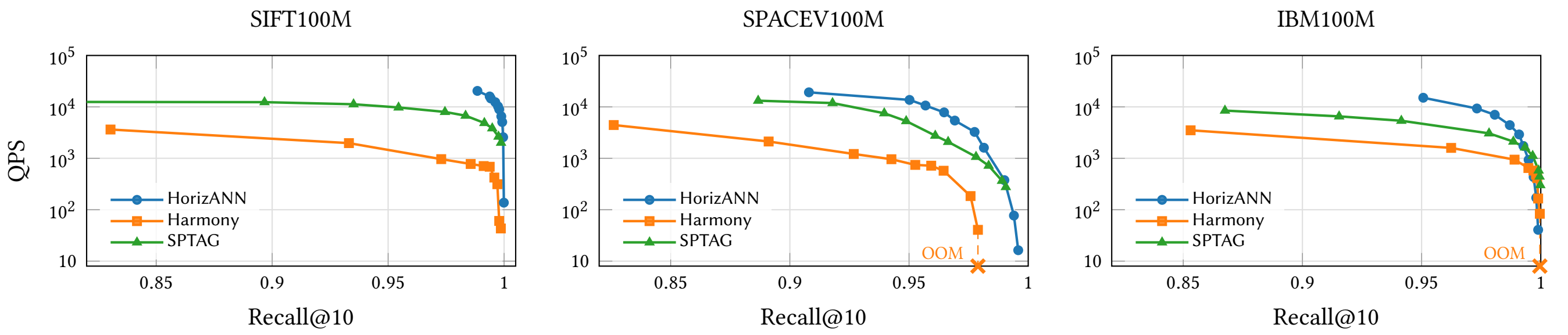
\begin{figure*}[t]
\centering
\begin{tikzpicture}
\pgfmathsetlengthmacro{\ylabw}{1.15cm}  
\pgfmathsetlengthmacro{\rpad}{0.1cm}    
\pgfmathsetlengthmacro{\hsep}{0.95cm}   
\pgfmathsetlengthmacro{\axw}{(\linewidth-\ylabw-\rpad-2*\hsep)/3}
\begin{groupplot}[
    group style={group size=3 by 1, horizontal sep=\hsep,
        xlabels at=edge bottom, ylabels at=edge left},  
    scale only axis,          
    width=\axw,
    height=2.4cm,             
    xlabel={Recall@10},
    ylabel={QPS},
    ymode=log,
    log basis y=10,
    ymin=8, ymax=100000,
    ytick={10,100,1000,10000,100000},
    yticklabels={$10$, $10^2$, $10^3$, $10^4$, $10^5$},
    xticklabel style={/pgf/number format/.cd, fixed, precision=3},
    scaled x ticks=false,
    grid=major,
    grid style={gray!25},
    tick label style={font=\scriptsize},
    label style={font=\small},
    title style={font=\small},
    every axis plot/.append style={mark size=1.2pt},
    legend pos=south west,
    legend cell align=left,
    legend style={draw=none, fill=white, fill opacity=0.8, text opacity=1,
        font=\scriptsize, nodes={inner sep=0.8pt}},
]
\nextgroupplot[title={SIFT100M}, xmin=0.82, xmax=1.005,
    xtick={0.85,0.9,0.95,1}]
\paretoplot[select=10, from=0.82]{horizann}{figures/exp/search_performance/horizann_sift100M.csv}
\addlegendentry{\system}
\paretoplot[y=server_qps]{harmony}{figures/exp/search_performance/harmony_sift100M.csv}
\addlegendentry{\textsf{Harmony}}
\paretoplot{sptag}{figures/exp/search_performance/sptag_sift100M.csv}
\addlegendentry{\textsf{SPTAG}}

\nextgroupplot[title={SPACEV100M}, xmin=0.82, xmax=1.00,
    xtick={0.85,0.9,0.95,1}]
\paretoplot[select=10, from=0.82]{horizann}{figures/exp/search_performance/horizann_spacev100M.csv}
\addlegendentry{\system}
\paretoplot[select=10, from=0.82, y=server_qps, oom=harmonycolor]{harmony}{figures/exp/search_performance/harmony_spacev100M.csv}
\addlegendentry{\textsf{Harmony}}
\paretoplot[select=10, from=0.82]{sptag}{figures/exp/search_performance/sptag_spacev100M.csv}
\addlegendentry{\textsf{SPTAG}}

\nextgroupplot[title={\gmm{100M}}, xmin=0.82, xmax=1.00,
    xtick={0.85,0.9,0.95,1}]
\paretoplot[select=10, from=0.82]{horizann}{figures/exp/search_performance/horizann_gmm100M.csv}
\addlegendentry{\system}
\paretoplot[select=10, from=0.82, y=server_qps, oom=harmonycolor]{harmony}{figures/exp/search_performance/harmony_gmm100M.csv}
\addlegendentry{\textsf{Harmony}}
\paretoplot[select=10, from=0.82]{sptag}{figures/exp/search_performance/sptag_gmm100M.csv}
\addlegendentry{\textsf{SPTAG}}
\end{groupplot}
\end{tikzpicture}
\vspace{-2mm}
\caption{Recall@10--QPS trade-off on 100M-scale datasets (because baselines failed on 1B). \system achieves the highest QPS on all three 100M-scale datasets up to a Recall@10 of 0.99, with 1.4--3.9$\times$ the QPS of \sptag and 5--27$\times$ that of \harmony at 0.99.}
\label{fig:eval:search_perf}
\end{figure*}

\subsection{Experimental Setup}
\label{sec:eval:setup}

\paragraph{Hardware.}
All experiments run on CloudLab~\cite{cloudlab} using c6320 nodes at the Clemson cluster, each with two 14-core Intel Xeon E5-2683 v3 CPUs (2.0\,GHz, 28 cores total), 256\,GiB RAM, and two 1\,TB SATA HDDs, connected via 10\,Gbps Ethernet. We use 12~machines for billion-scale datasets.

\paragraph{Datasets.}
Table~\ref{tab:datasets} presents the datasets. SIFT~\cite{jégou2011searchingbillionvectorsrerank} and SPACEV~\cite{simhadri2022resultsneurips21challengebillionscale} are standard billion-scale benchmarks. We further generate \gmm{1B}, a synthetic dataset of 1~billion 128-dimensional vectors sampled from a Gaussian Mixture Model with 300 components which learnt from real world embeddings.

\paragraph{Baselines.}
We compare against two distributed vector search systems.
\textsf{SPTAG}~\cite{sptag} is Microsoft’s ANN library; we use its SPFresh~\cite{spfresh} index, which organizes vectors into posting lists via SPANN-style~\cite{spann} hierarchical balanced clustering and supports in-place updates. For a fair comparison, we place \textsf{SPTAG}’s on-disk index on a 180–200 GiB RAM disk (sized per dataset) on each worker, so that its entire index resides in memory.
\textsf{Harmony}~\cite{harmony} partitions data using a hybrid of dimension-based and vector-based strategies with adaptive query routing.

\paragraph{Hyperparameters.}
All workers build HNSW indexes with $M{=}32$ and $\mathit{ef}_{\text{con}}{=}200$. We vary $\mathit{ef}_{\text{search}}$ (default 500) and routing fan-out $k_{\text{seed}}$ (default 10) to trace QPS--recall curves. We report Recall@10 and throughput in queries per second (QPS), measuring over 10K queries with brute-force ground truth.

\subsection{Index Build Time}
\label{sec:eval:build}



This experiment aims to show that \system efficiently indexes a growing dataset incrementally as the cluster expands, which existing systems fail to do.

\paragraph{Setup.}
We grow each index from 0 to 1B vectors in ten phases of 100M vectors each. \system starts with a single worker; after each phase except the last, we add one worker and trigger repartitioning, so the cluster grows from 1 to 10 workers. The reported insertion time includes repartitioning time. In addition, we use batch insertion with a batch size of 5K.
For \harmony, as it does not support incremental insertion, we conduct a bulk build at each phase with the same number of workers as \system.
For \sptag, we extend it so that each worker builds an initial index on half of its shard and incrementally inserts the other half in batches via SPFresh’s in-place updates. The reported insertion time includes SPFresh’s background split time. We disable the reassignment for faster insertion and insert all 100M vectors in one phase to a specific worker with a batch size of 10K.
For each phase, \cref{fig:eval:incremental_build_time} reports the time to insert the next 100M vectors.
We cap the time to build each full index at 24 hours.

\paragraph{Results.}
\system inserts each 100M vectors in 49--148 minutes across the three datasets.
More importantly, its per-phase time does not grow with the dataset size and worker counts. In contrast, both baselines fail to reach 1B vectors. \sptag exceeds the 24-hour limit while inserting 300M vectors for SIFT1B and SPACEV1B, and 100M vectors for \gmm{1B}. \harmony runs out of memory at 300M vectors on SIFT1B and \gmm{1B}, and 400M on SPACEV1B.

\paragraph{Analysis.}
\system keeps its per-phase time nearly consistent because each worker will hold only about 100 million vectors throughout the entire insertion process. For example, after the last repartitioning on \gmm{1B}, worker sizes range from 78M to 104M. 
Each new 100M-vector block is spread over all current workers, so the insertion work per batch shrinks as the cluster grows, from 240\,ms to 58\,ms per 5K-vector batch on \gmm{1B} between the first and the last phase. 
\sptag cannot move vectors between shards, so a single worker must absorb each 100M-vector block while the other workers stay idle. On a single worker, SPFresh slows down as the shard grows. Building the initial index over the first 50M vectors already takes 7.8--8.6 hours, the time per 1M inserted vectors then rises from 88\,s to 1{,}074\,s on \gmm{1B} as more vectors are inserted, waiting for background splits adds another 8.5 hours, so insertion on \gmm{1B} reaches 24-hour cap in the first phase.
\harmony has no incremental path, so each phase rebuilds the index over all vectors inserted so far. The build is dominated by assigning each vector to its nearest centroid (80--86\% of the build time), which compares every vector against all $\sqrt{N}$ centroids and thus grows as $N^{1.5}$; it runs on the main node alone, so it does not shrink as workers are added. 
Its main node also runs out of memory once the dataset reaches about 40B vector components (300M 128-dimensional vectors from SIFT1B and \gmm{1B}, and 400M 100-dimensional vectors from SPACEV1B), regardless of the number of workers.

\subsection{Search Performance}
\label{sec:eval:search}


This experiment aims to show that \system achieves higher search performance than existing distributed systems, despite its support for elasticity.

\paragraph{Setup.}
Because \harmony fails to build an index on any of the three 1B datasets, we evaluate search performance on the first 100M vectors of each dataset with 10 workers.
For \system, we perform queries in batches of 5K on the incrementally built index, and each worker searches in parallel on all its cores. We vary $\mathit{ef}_{\text{search}}$ and $k_{\text{seed}}$.
For \sptag, we bulk-build each shard's index over its 10M vectors. Its client sends queries in 64 threads, and the aggregator forwards each query to every shard which serves requests with 28 threads. We vary \texttt{SearchInternalResultNum}, the number of postings each query searches on each shard.
\harmony builds an IVF index with $\sqrt{N} = 10{,}000$ lists, takes the whole query set as one batch, and pipelines it in groups of 2{,}000 queries through 5 teams of 2 workers, each using all its cores. We vary $n_{\text{probe}}$ only.
\cref{fig:eval:search_perf} presents the Recall@10--QPS curves.

\paragraph{Results.}
Up to a Recall@10 of 0.99, \system achieves the highest Recall@10 at a given QPS, or equivalently, the highest QPS at a given Recall@10.
Across the three datasets, at Recall@10 = 0.95, \system delivers 2.1--3.9$\times$ the QPS of \sptag and 9--21$\times$ that of \harmony.
At 0.99, its advantage is 1.4--3.9$\times$ over \sptag and 5--27$\times$ over \harmony on SIFT100M and \gmm{100M}, while \harmony cannot reach 0.98 on SPACEV100M before running out of memory.

\paragraph{Analysis.}
First, as demonstrated in \cref{sec:index-query}, \system partitions data with balanced k-means, so each worker holds a cluster of nearby vectors, and the main node routes each query only to the few partitions that hold its nearest routing samples. 
In contrast, \sptag broadcasts every query to all workers, therefore every query consumes work on every machine. 
Second, within a partition \system runs an in-memory graph search whose cost grows slowly with the search budget, whereas \sptag and \harmony scan entire posting lists or IVF lists, whose cost grows linearly with \texttt{SearchInternalResultNum} or $n_{\text{probe}}$. 
Across the three datasets, raising $n_{\text{probe}}$ from 10 to 100 cuts \harmony's QPS by 5.1--6.1$\times$, and raising \texttt{SearchInternalResultNum} from 8 to 96 cuts \sptag's QPS by 2.4--3.1$\times$, while \system already exceeds 0.95 recall at its smallest $\mathit{ef}_{\text{search}}$ of 500 on all three datasets.

\begin{figure}[t]
\centering
\captionsetup[sub]{skip=1pt}
\pgfmathsetlengthmacro{\qpsw}{\columnwidth-1.5cm}
\def\siftsweepcsv{figures/exp/scalability/horizann_sweep.csv}

\begin{subfigure}{\columnwidth}
\centering
\begin{tikzpicture}[trim axis left, trim axis right]
\begin{axis}[
    scale only axis,
    width=\qpsw,
    height=2.5cm,
    xlabel={Indexed vectors (M)},
    ylabel={QPS},
    xmin=0, xmax=1050,
    xtick={100,200,...,1000},
    /pgf/number format/1000 sep={},
    ymin=0, ymax=30000,
    ytick={0, 10000, 20000, 30000},
    grid=major,
    grid style={gray!25},
    tick label style={font=\small},
    label style={font=\small},
    legend pos=north west,
    legend cell align=left,
    legend style={draw=none, fill=white, fill opacity=0.8, text opacity=1,
        font=\small, nodes={inner sep=0.8pt}},
]
\addplot[dataset sift] table[
    col sep=comma,
    x expr=\thisrow{total_inserted}/1000000,
    y=qps,
] {figures/exp/index_build/horizann_sift1b.csv};
\addlegendentry{SIFT1B}
\addplot[dataset spacev] table[
    col sep=comma,
    x expr=\thisrow{total_inserted}/1000000,
    y=qps,
] {figures/exp/index_build/horizann_spacev1b.csv};
\addlegendentry{SPACEV1B}
\addplot[dataset gmm] table[
    col sep=comma,
    x expr=\thisrow{total_inserted}/1000000,
    y=qps,
] {figures/exp/index_build/horizann_gmm1b.csv};
\addlegendentry{\gmm{1B}}
\end{axis}
\end{tikzpicture}
\caption{QPS after each 100M-vector block is inserted.}
\label{fig:eval:incremental_build_qps:scale}
\end{subfigure}

\vspace{4mm}

\begin{subfigure}{\columnwidth}
\centering
\begin{tikzpicture}[trim axis left, trim axis right]
\begin{axis}[
    scale only axis,
    width=\qpsw,
    height=2.5cm,
    xlabel={Recall@10},
    ylabel={QPS},
    ylabel near ticks,
    ylabel shift=-2pt,
    xmin=0.985, xmax=1.0,
    xtick={0.985,0.99,0.995,1},
    xticklabel style={/pgf/number format/.cd, fixed, fixed zerofill, precision=3},
    scaled x ticks=false,
    ymode=log, log basis y=10,
    ymin=100, ymax=100000,
    ytick={100,1000,10000,100000},
    grid=major,
    grid style={gray!25},
    tick label style={font=\small},
    label style={font=\small},
    legend pos=south west,
    legend columns=2,
    legend cell align=left,
    legend style={draw=none, fill=white,
        font=\small, nodes={inner sep=0.8pt},
        /tikz/every even column/.append style={column sep=0.6em}},
    every axis plot/.append style={thick, mark=*, mark size=1.0pt,
        mark options={solid}},
]
\paretoplot[x=recall, from=0.985, select=10,
    where=total_inserted, equals=100000000]
    {siftcolor!40, densely dotted}{\siftsweepcsv}
\addlegendentry{100M}
\paretoplot[x=recall, from=0.985, select=10,
    where=total_inserted, equals=200000000]
    {siftcolor!60, dashdotted}{\siftsweepcsv}
\addlegendentry{200M}
\paretoplot[x=recall, from=0.985, select=10,
    where=total_inserted, equals=500000000]
    {siftcolor!80, dashed}{\siftsweepcsv}
\addlegendentry{500M}
\paretoplot[x=recall, from=0.985, select=10,
    where=total_inserted, equals=1000000000]
    {siftcolor, solid}{\siftsweepcsv}
\addlegendentry{1B}
\end{axis}
\end{tikzpicture}
\caption{Recall@10--QPS curves on SIFT as the index grows.}
\label{fig:eval:incremental_build_qps:recall}
\end{subfigure}

\vspace{-2mm}
\caption{Scalability of \system as the index grows from 100M to 1B vectors and the cluster from 1 to 10 workers.
(a)~Query throughput increases 2.4--3.0$\times$ on the three datasets.
(b)~Recall@10 on SIFT1B stays consistent.}
\label{fig:eval:incremental_build_qps}
\end{figure}
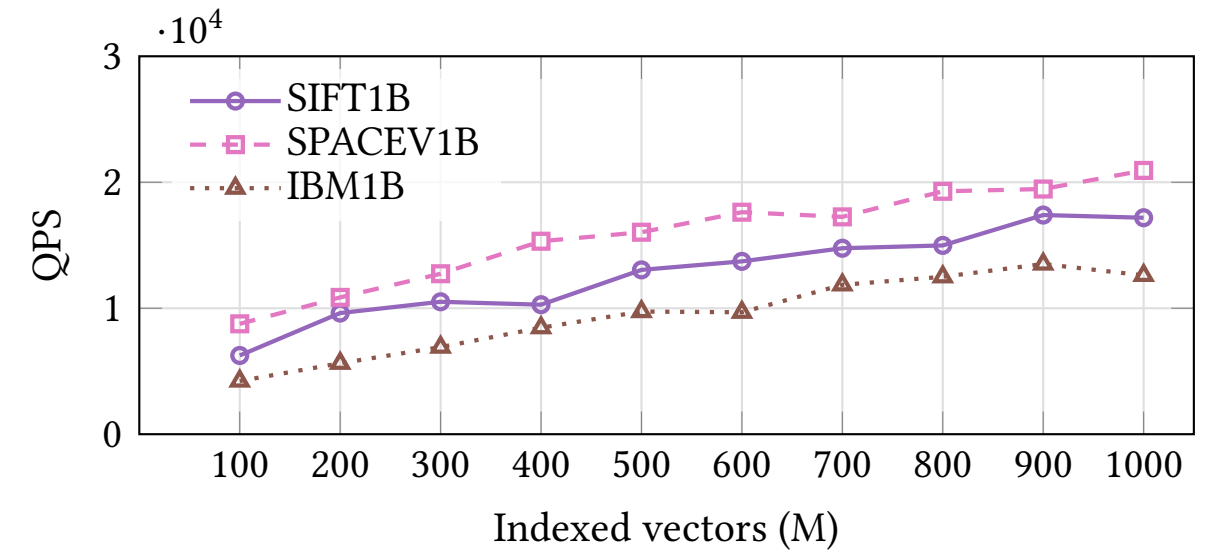
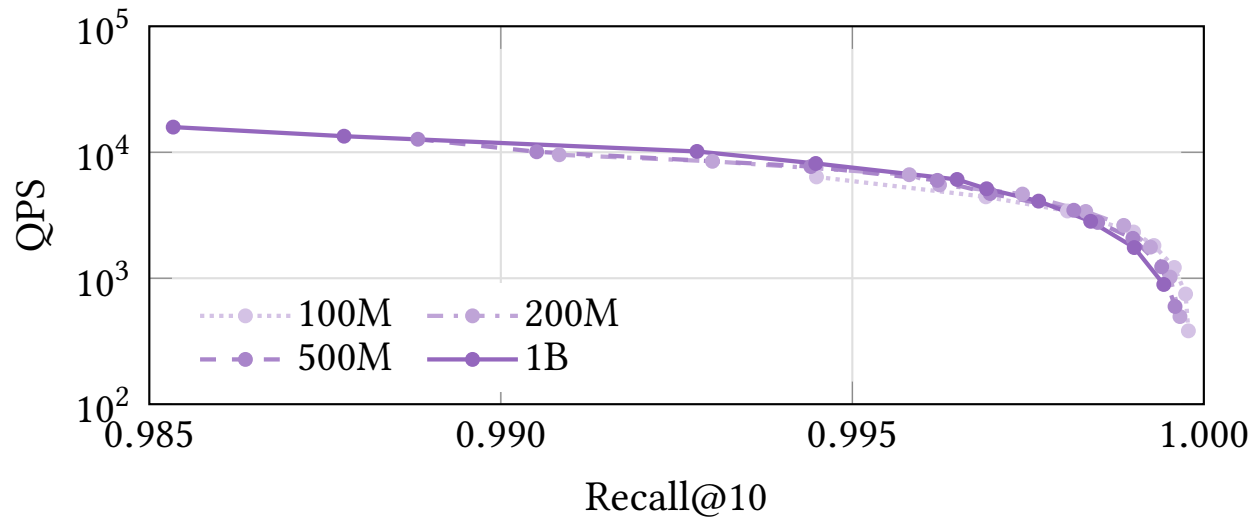

\subsection{Scalability}
\label{sec:eval:scalability}


This experiment aims to show that \system's query throughput grows as workers are added, while its recall is preserved.

\paragraph{Setup.}
We apply the same incremental index building setup as \cref{sec:eval:build}. After the insertion of each phase ends, we run 10K queries to measure the throughput as QPS with default parameters. And we perform the additional search sweep at 100M, 200M, 500M and 1B vectors to measure the Recall@10--QPS curve in the same way as \cref{sec:eval:search}. \cref{fig:eval:incremental_build_qps} reports the details.

\paragraph{Results.}
The QPS increases 2.4 -- 3.0$\times$ from 100M to 1B vectors across three datasets (\cref{fig:eval:incremental_build_qps:scale}), which shows \system can easily scale up to larger datasets and clusters. Four curves nearly overlap with each other (\cref{fig:eval:incremental_build_qps:recall}), which means \system's search performance won't change as cluster and index sizes change.

\paragraph{Analysis.}
Because each query is routed to only a few workers and each worker's index stays at about 100M vectors from 1 to 10 workers, each batch of 5K queries is spread over more workers at the same cost per query.
As a result, the time the workers spend on a batch falls by 3.4--3.6$\times$, which means the busiest worker's share drops from all 5K queries to about 1.4--1.5K.
End to end, a batch takes 0.56--1.18\,s on 1 worker and 0.23--0.39\,s on 10 workers. This is less than 3.4--3.6$\times$ because each batch also spends 66--90\,ms on routing, merging results and client communication. This part takes 6--14\% of the batch time at 100M but 21--38\% at 1B.
In addition, as the cluster grows, the nearest neighbors of each query spread over more workers. To keep a comparable recall, a larger cluster needs a higher $k_{\text{seed}}$ and $\mathit{ef}_{\text{search}}$, which costs QPS. As analyzed above, the extra workers absorb this work, so the four curves in \cref{fig:eval:incremental_build_qps:recall} nearly overlap at high recall: at Recall@10 of 0.998, QPS stays at 3.4--3.5K from 100M to 1B.


\ExplSyntaxOn
\cs_set_eq:NN \rbfp \fp_eval:n
\ExplSyntaxOff
\begin{figure}[t]
\centering
\pgfplotstableread[col sep=comma]{figures/exp/rebalancing/horizann_insertion.csv}\rbinsert

\pgfplotstablegetrowsof{\rbinsert}
\pgfmathtruncatemacro\rbnlast{\pgfplotsretval-1}
\def\rbclock{0}\def\rbrebtotal{0}
\def\rbline{(0,0)}\def\rbwins{}\def\rbprev{}
\pgfplotsforeachungrouped \rbi in {0,...,\rbnlast}{%
    \pgfplotstablegetelem{\rbi}{phase}\of\rbinsert \let\rbphase\pgfplotsretval
    \ifx\rbphase\rbprev\else
        \let\rbprev\rbphase
        \pgfplotstablegetelem{\rbi}{total_inserted}\of\rbinsert \let\rbn\pgfplotsretval
        \pgfplotstablegetelem{\rbi}{insert_sec}\of\rbinsert     \let\rbins\pgfplotsretval
        \pgfplotstablegetelem{\rbi}{add_worker_ms}\of\rbinsert  \let\rbreb\pgfplotsretval
        \edef\rbs{\rbfp{\rbclock+\rbins/60}}
        \edef\rbe{\rbfp{\rbs+\rbreb/60000}}%
        \edef\rby{\rbfp{\rbn/1e6}}%
        \edef\rbline{\rbline (\rbs,\rby) (\rbe,\rby)}%
        \edef\rbwins{\unexpanded\expandafter{\rbwins}\noexpand\rbwin{\rbs}{\rbe}{\rby}%
            {\rbfp{\rbreb/60000}}{\rbfp{100*\rbreb/(\rbreb+1000*\rbins)}}}%
        \let\rbclock\rbe
        \edef\rbrebtotal{\rbfp{\rbrebtotal+\rbreb/60000}}%
    \fi}
\edef\rbrebshare{\rbfp{100*\rbrebtotal/\rbclock}}

\newcommand{\rbymax}{600}   
\pgfmathsetlengthmacro{\rbw}{\linewidth-1.5cm}  
\pgfplotsset{
    rbcommon/.style={
        scale only axis, width=\rbw, height=2.3cm,
        ylabel near ticks,  
        every axis y label/.append style={inner sep=0pt},     
        every y tick label/.append style={inner sep=1.5pt},
        tick label style={font=\small},
        label style={font=\small},
        every axis x label/.append style={text height=1.6ex, text depth=0.5ex},
        xlabel shift=-2pt,  
        every x tick label/.append style={text height=1.2ex, text depth=0.25ex},
        grid=major, grid style={gray!25},
        set layers,  
    },
}
\tikzset{
    rb window/.style={fill=rebalcolor!25, draw=rebalcolor!25,  
        /pgfplots/on layer=axis background},
    rb label/.style={font=\footnotesize, text=rebalcolor!80!black, inner sep=1.5pt},
}
\newcommand{\rbwin}[5]{%
    \addplot[rb window, forget plot] coordinates {(#1,0) (#1,#3) (#2,#3) (#2,0)} -- cycle;
    \node[rb label, anchor=south east] at (axis cs:#2,#3)
        {\pgfmathprintnumber[fixed, fixed zerofill, precision=1]{#4}\,min
         (\pgfmathprintnumber[fixed, precision=0]{#5}\%)};}

\begin{tikzpicture}
\begin{axis}[
    rbcommon,
    xlabel={Time (min)},
    ylabel={Inserted vectors (M)},
    xmin=0, xmax=\rbclock,
    xtick={0,60,...,300},
    minor xtick={30,90,...,270},
    ymin=0, ymax=\rbymax,
    ytick={0,100,...,600},
    xmajorgrids=false,  
    clip=false,
]
\rbwins
\edef\rbtmp{\noexpand\addplot[insertcolor, thick] coordinates {\rbline};}\rbtmp
\node[rb label, anchor=north west, fill=white] at (rel axis cs:0.01,0.97)
    {Total repartitioning:
     \pgfmathprintnumber[fixed, fixed zerofill, precision=1]{\rbrebtotal}\,min
     (\pgfmathprintnumber[fixed, fixed zerofill, precision=1]{\rbrebshare}\%)};
\end{axis}
\end{tikzpicture}
\vspace{-3mm}
\caption{Insertion and repartitioning of \system on the first 500M vectors of
SIFT1B. Each bar is one repartitioning, labelled with its duration and its share
of the phase. Each repartitioning finishes within 14 minutes, and together they take 15\% of the total build time to 500M vectors.}
\label{fig:eval:rebalancing:timeline}
\end{figure}
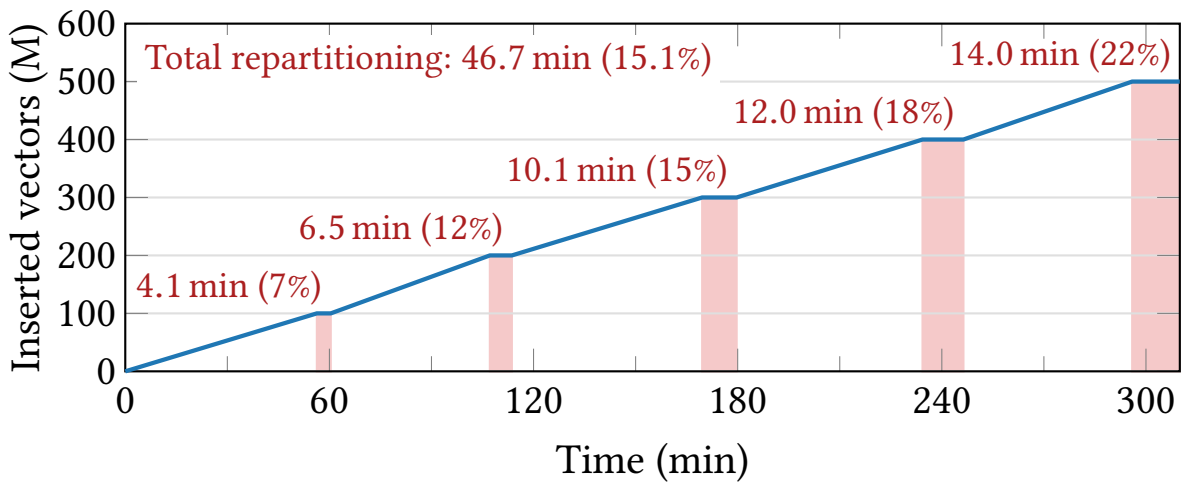

\subsection{Repartitioning Efficiency}
\label{sec:eval:rebalancing_efficiency}



This experiment aims to show that \system repartitions its index quickly, at a small cost relative to index building.

\begin{table}[t]
\centering
\caption{Repartitioning time on SIFT by phase, in seconds. Planning on the main node (sampling and balanced $k$-means, 1--4\,s) is not included. 
Transfer \& commit takes 53--64\% of each repartitioning, and every stage grows with the index as ghost edges accumulate.
$^\dagger$Measured while queries ran continuously (discussed in \cref{sec:eval:rebalancing_liveness}).}
\label{tab:rebalance-breakdown}
\vspace{-2mm}
\small
\begin{tabular}{rrrrr}
\toprule
Vectors & Prepare & Transfer \& commit   & Retire \& persist     & Total (s) \\
\midrule
100M           & 66.7  & 156.1 & 21.5  & 244.3 \\
200M           & 95.8  & 243.4 & 48.2  & 387.4 \\
300M           & 139.3 & 388.2 & 78.7  & 606.2 \\
400M           & 166.8 & 460.0 & 91.5  & 718.2 \\
500M$^\dagger$ & 254.7 & 446.4 & 135.0 & 836.1 \\
\bottomrule
\end{tabular}
\end{table}

\paragraph{Setup.}
We apply the same incremental index building setup as in \cref{sec:eval:build}. We build the index on SIFT1B up to 500M vectors (using up to 6 worker nodes) and record the repartitioning time for each 100M vector phase.

\paragraph{Results.}
\cref{fig:eval:rebalancing:timeline} presents the absolute time of repartitioning in minutes and its share of each phase. The repartitioning time grows from 4 minutes (100M vectors) to 14 minutes (500M vectors), and it takes from 7\% of total insertion time to 22\% of total insertion time. \cref{tab:rebalance-breakdown} shows the stage-wise time consumption of repartitioning in each phase. \emph{Transfer \& commit} accounts for the largest share of repartitioning time, during which workers stream split subindexes to one another and merge incoming subindexes.

\paragraph{Analysis.}
Repartitioning time grows with the dataset size, mainly because the number of ghost edges grows. 
As described in \cref{sec:index-insert-delete}, \system inserts each vector into its resident worker, and uses the routing index to find other workers that may hold its nearest neighbors. It searches these workers for the vector's nearest neighbors and builds ghost edges, which are stored on its resident worker. Therefore, the number of vectors with ghost edges on each worker keeps increasing.
On the first worker, the share of vectors with ghost edges grows from 14\% after the repartitioning at 100M to 59\% after the one at 500M, and its ghost store grows from 0.5 to 17\,GiB. As a result, each worker spends more time loading ghost edges from disk (\emph{Prepare}), activating the ghost edges of each vector while merging indexes (line~\ref{alg:hnsw-repartition:reactivate} in \cref{alg:hnsw-repartition}, \emph{Transfer \& commit}), and persisting them back to disk (\emph{Retire \& persist}).

\ExplSyntaxOn
\cs_set_eq:NN \rbfp \fp_eval:n
\ExplSyntaxOff
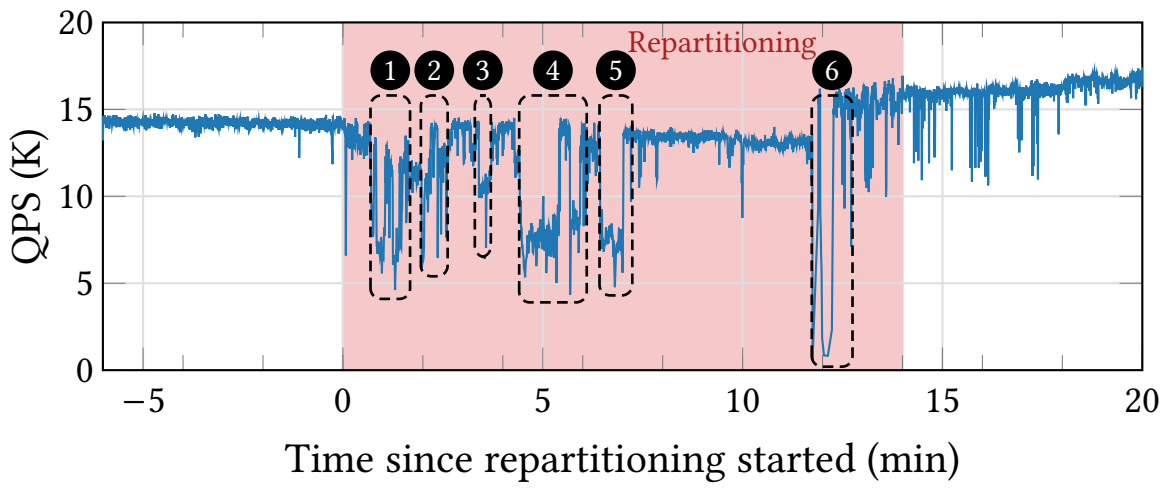
\begin{figure}[t]
\centering
\pgfplotstableread[col sep=comma]{figures/exp/rebalancing/horizann_insertion.csv}\rbinsert

\pgfplotstablegetrowsof{\rbinsert}
\pgfmathtruncatemacro\rbnlast{\pgfplotsretval-1}
\pgfplotstablegetelem{\rbnlast}{add_worker_start_unix_ms}\of\rbinsert \let\rbt\pgfplotsretval
\pgfplotstablegetelem{\rbnlast}{add_worker_end_unix_ms}\of\rbinsert
\edef\rbdur{\rbfp{(\pgfplotsretval-\rbt)/60000}}  

\newcommand{\rbqpsmax}{20}  
\pgfmathsetlengthmacro{\rbgap}{0.8cm}  
\pgfmathsetlengthmacro{\rbw}{\linewidth-2*\rbgap}
\pgfplotsset{
    rbcommon/.style={
        scale only axis, width=\rbw, height=2.3cm,
        ylabel near ticks,  
        every axis y label/.append style={inner sep=0pt},     
        every y tick label/.append style={inner sep=1.5pt},
        tick label style={font=\footnotesize},
        label style={font=\small},
        every axis x label/.append style={text height=1.6ex, text depth=0.5ex},
        xlabel shift=-2pt,  
        every x tick label/.append style={text height=1.2ex, text depth=0.25ex},
        grid=major, grid style={gray!25},
        set layers,  
    },
}
\tikzset{
    rb window/.style={fill=rebalcolor!25, draw=rebalcolor!25,  
        /pgfplots/on layer=axis background},
    rb label/.style={font=\scriptsize, text=rebalcolor!80!black, inner sep=1.5pt},
    rb dipnum/.style={circle, fill=black, text=white, font=\scriptsize, inner sep=1pt, outer sep=1pt},
}
\newcommand{\rbdiptop}{15.8}  
\newcommand{\rbdip}[4]{
    \draw[line width=0.5pt, dash pattern=on 2pt off 1.2pt, rounded corners=2pt]
        (axis cs:#1,#3) rectangle (axis cs:#2,\rbdiptop);
    \node[rb dipnum, above] at (axis cs:{(#1+#2)/2},\rbdiptop) {#4};}

\begin{tikzpicture}[trim axis left, trim axis right]
\begin{axis}[
    rbcommon,
    xlabel={Time since repartitioning started (min)},
    ylabel={QPS (K)},
    enlarge x limits=false,
    xtick={-5,0,...,20},
    minor xtick={-6,-4,...,20},
    ymin=0, ymax=\rbqpsmax,
    ytick={0,5,...,\rbqpsmax},
]
\addplot[rb window, forget plot] coordinates {
    (0,0) (0,\rbqpsmax) (\rbdur,\rbqpsmax) (\rbdur,0)} -- cycle;
\node[rb label, anchor=north] at (axis cs:9.5,\rbqpsmax) {Repartitioning};  
\addplot[horizanncolor, line width=0.4pt]
    table[col sep=comma, x=t_min, y=qps_k] {figures/exp/rebalancing/horizann_rebalancing_liveness_plot.csv};
\rbdip{0.69}{1.68}{4.1}{1}
\rbdip{1.95}{2.62}{5.4}{2}
\rbdip{3.30}{3.70}{6.5}{3}
\rbdip{4.41}{6.10}{3.9}{4}
\rbdip{6.42}{7.24}{4.3}{5}
\rbdip{11.73}{12.75}{0.2}{6}
\end{axis}
\end{tikzpicture}
\vspace{-2mm}
\caption{Query throughput of \system around the repartitioning at 500M vectors of
SIFT1B, from 6 minutes before it starts to 6 minutes after it ends.
Boxes~\mbox{\protect\circled{1}--\protect\circled{6}} mark the throughput
valleys. On average, \system retains 84\% of its pre-repartitioning QPS while repartitioning.}
\label{fig:eval:rebalancing:qps}
\end{figure}

\paragraph{Limitation.}
Repartitioning takes about 2.5 more minutes with every 100M vectors, while the insertion time of a phase stays flat (\cref{sec:eval:build}).
With far more vectors or workers (e.g., 10B vectors or 50 workers), repartitioning will eventually take longer than insertion in some phase.
Dropping ghost edges that are unlikely to be reactivated could bound this cost, which we leave to future work.

\subsection{Online Repartitioning}
\label{sec:eval:rebalancing_liveness}


This experiment aims to show that \system sustains high query throughput and recall while repartitioning.

\paragraph{Setup.}
We apply the same setup as in \cref{sec:eval:rebalancing_efficiency}. During the repartitioning phase at 500M vectors, we continuously issue queries from a client and record the QPS prior to repartitioning for 6 minutes. Next, we initiate the repartitioning process to add the sixth worker, while maintaining the ongoing query operations from the client. Once repartitioning is complete, we conduct another 6-minute period to measure the QPS following the repartitioning (all measured under default parameters).

\paragraph{Results.}
\cref{fig:eval:rebalancing:qps} shows the query throughput before, during, and after repartitioning on 500M vectors.
\system answer queries during repartitioning with 11.7K QPS on average (84\% of the 14.0K QPS before repartitioning), while Recall@10 stays at 0.98. And there are several QPS drops (marked as \mbox{\circled{1}--\circled{6}}). 

\paragraph{Analysis.}
Search and repartitioning share the CPU cores of each worker, so throughput drops whenever workers run a compute-heavy repartitioning step.
In \circled{1}, all five workers split their index at the same time.
In \circled{2} and \circled{3}, workers copy the vectors they keep into a new index, four workers at once in \circled{2}, and the last one in \circled{3}.
In \circled{4}, transfer starts and every worker merges the vectors it receives at the same time.
In \circled{5}, two workers merge the largest pieces they receive.
Outside these drops, throughput stays at 93--94\% of its pre-repartitioning level: loading and saving ghost edges are primarily disk-bound rather than CPU-bound, and after \circled{5}, only the new worker continues merging and does not yet serve queries.
Finally, \circled{6} corresponds to the main node switching to the new routing index, so throughput falls below 1K QPS for a moment right after the switch. Subsequently, as worker nodes release the old index, throughput drops again.

\subsection{Memory Usage}
\label{sec:eval:memory}

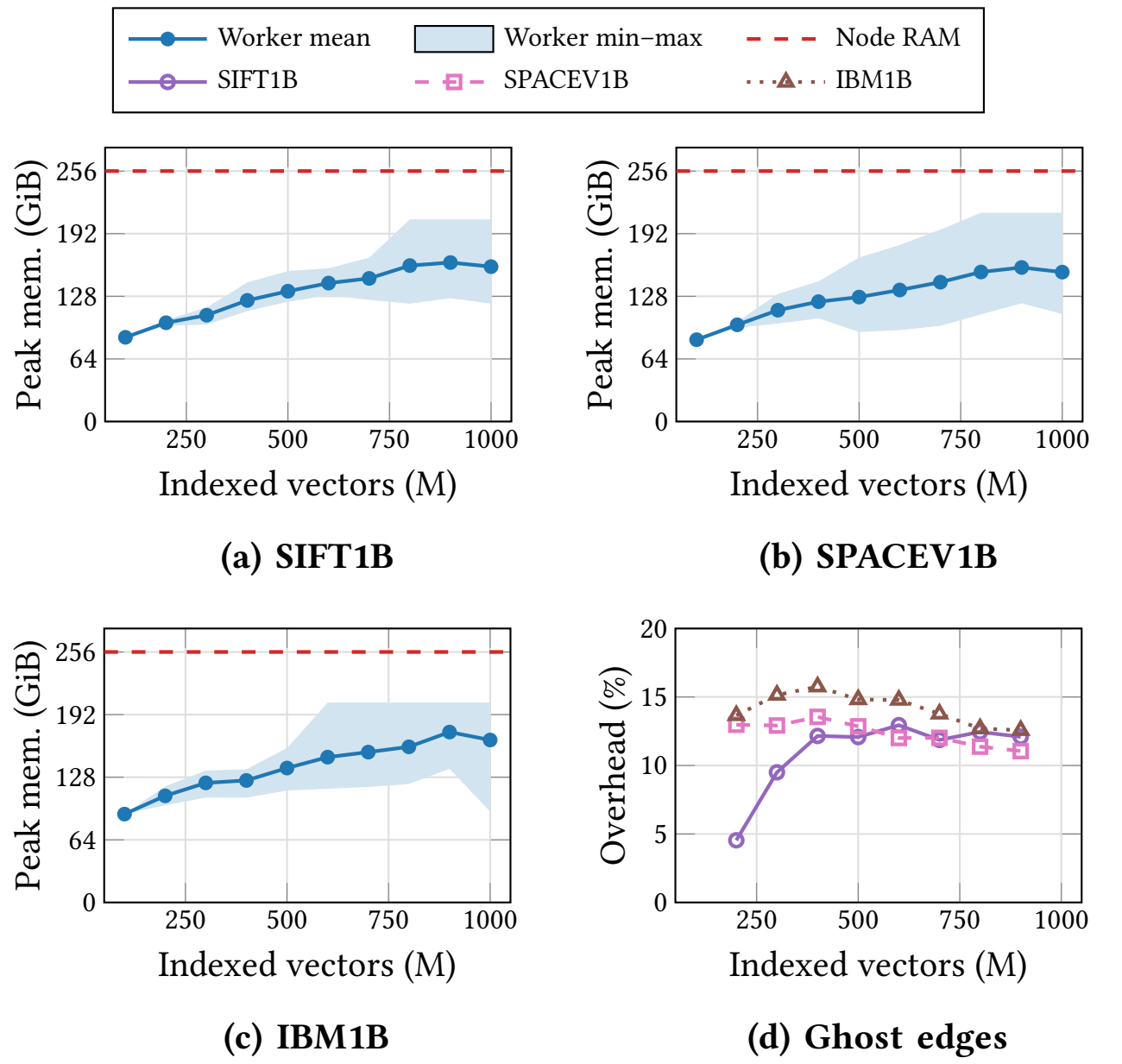
\begin{figure}[t]
\centering
\captionsetup[sub]{skip=0pt}  
\pgfplotstableread[col sep=comma]{figures/exp/memory/horizann_sift1b.csv}\ghostsift
\pgfplotstableread[col sep=comma]{figures/exp/memory/horizann_spacev1b.csv}\ghostspacev
\pgfplotstableread[col sep=comma]{figures/exp/memory/horizann_gmm1b.csv}\ghostgmm
\pgfplotstableread[col sep=comma]{figures/exp/memory/horizann_sift1b_worker_summary.csv}\wsift
\pgfplotstableread[col sep=comma]{figures/exp/memory/horizann_spacev1b_worker_summary.csv}\wspacev
\pgfplotstableread[col sep=comma]{figures/exp/memory/horizann_gmm1b_worker_summary.csv}\wgmm
\pgfmathsetlengthmacro{\axw}{3cm}  
\pgfplotsset{
    memcommon/.style={
        scale only axis, width=\axw, height=2.1cm,
        xlabel={Indexed vectors (M)},
        xmin=50, xmax=1050,
        xtick={250,500,750,1000},
        /pgf/number format/1000 sep={},
        grid=major, grid style={gray!25},
        tick label style={font=\scriptsize, inner sep=1.5pt},  
        label style={font=\small},
        ylabel shift={-3pt},                                    
    },
    memworker/.style={
        memcommon,
        ymin=0, ymax=280,
        ytick={0,64,128,192,256},
    },
    mem mean/.style={horizanncolor, thick, mark=*, mark size=1.2pt},
    mem band/.style={fill=horizanncolor!20, draw=none},
}
\tikzset{
    mem limit/.style={limitcolor, dashed, thick},  
    mem subcap/.style={anchor=north, text width=\axw, align=center,
        inner sep=0pt, yshift=-4pt},
}
\newcommand{\workermem}[1]{%
    \addplot[draw=none, name path=lo, forget plot] table[
        x expr=(\thisrow{phase}+1)*100, y expr=\thisrow{peak_min_mb}/1024] {#1};
    \addplot[draw=none, name path=hi, forget plot] table[
        x expr=(\thisrow{phase}+1)*100, y expr=\thisrow{peak_max_mb}/1024] {#1};
    \addplot[mem band, forget plot] fill between[of=lo and hi];
    \addplot[mem mean, forget plot] table[
        x expr=(\thisrow{phase}+1)*100, y expr=\thisrow{peak_mean_mb}/1024] {#1};
    \draw[mem limit] (axis cs:50,256) -- (axis cs:1050,256);
}
\newcommand{\measurehang}[3]{%
    \sbox0{\begin{tikzpicture}
        \begin{axis}[#1]\end{axis}
        \pgfpointdiff{\pgfpointanchor{current axis}{outer west}}{\pgfpointanchor{current axis}{west}}%
        \pgfgetlastxy{\hangx}{\hangy}\global\let#2\hangx
        \pgfpointdiff{\pgfpointanchor{current axis}{east}}{\pgfpointanchor{current axis}{outer east}}%
        \pgfgetlastxy{\hangx}{\hangy}\global\let#3\hangx
    \end{tikzpicture}}%
}
\measurehang{memworker, ylabel={Peak mem. (GiB)}}{\hangLa}{\hangR}   
\measurehang{memworker, ylabel={Peak mem. (GiB)}}{\hangLb}{\hangdummy} 
\measurehang{memcommon, ylabel={Overhead (\%)},
    ymin=0, ymax=20, ytick={0,5,10,15,20}}{\hangLd}{\hangdummy}      
\pgfmathsetlengthmacro{\gsep}{\hangR + max(\hangLb, \hangLd) + 0.35cm}
\pgfmathsetlengthmacro{\axw}{(\linewidth-\hangLa-\gsep-\hangR)/2 - 1pt}

\ref{mem-legend}\par\vspace{6pt}

\begin{tikzpicture}
\begin{axis}[memworker, name=a, ylabel={Peak mem. (GiB)}]
\workermem{\wsift}
\end{axis}
\node[mem subcap] (a-cap) at (a.south |- a.outer south)
    {\subcaption{SIFT1B}\label{fig:eval:memory:sift}};

\coordinate (b-pos) at ([xshift=\gsep]a.south east);
\begin{axis}[memworker, name=b, ylabel={Peak mem. (GiB)}, at={(b-pos)}, anchor=south west]
\workermem{\wspacev}
\end{axis}
\node[mem subcap] (b-cap) at (b.south |- b.outer south)
    {\subcaption{SPACEV1B}\label{fig:eval:memory:spacev}};

\coordinate (c-pos) at ([yshift=-10pt]a-cap.south west);
\begin{axis}[memworker, name=c, ylabel={Peak mem. (GiB)},
    at={(c-pos)}, anchor=north west]
\workermem{\wgmm}
\end{axis}
\node[mem subcap] (c-cap) at (c.south |- c.outer south)
    {\subcaption{\gmm{1B}}\label{fig:eval:memory:gmm}};

\coordinate (d-pos) at ([xshift=\gsep]c.south east);
\begin{axis}[
    memcommon, name=d, at={(d-pos)}, anchor=south west,
    ylabel={Overhead (\%)},
    ymin=0, ymax=20, ytick={0,5,10,15,20},
    legend to name={mem-legend},
    legend columns=3,
    legend style={font=\scriptsize, cells={anchor=west},
        /tikz/every even column/.append style={column sep=0.8em}},
]
\addlegendimage{mem mean}\addlegendentry{Worker mean}
\addlegendimage{area legend, mem band}\addlegendentry{Worker min--max}
\addlegendimage{mem limit}\addlegendentry{Node RAM}
\addplot[dataset sift] table[
    x expr=\thisrow{total_inserted}/1000000,
    y expr=100*\thisrow{ghost_memory_mb}/\thisrow{rss_before_mb},
] {\ghostsift};
\addlegendentry{SIFT1B}
\addplot[dataset spacev] table[
    x expr=\thisrow{total_inserted}/1000000,
    y expr=100*\thisrow{ghost_memory_mb}/\thisrow{rss_before_mb},
] {\ghostspacev};
\addlegendentry{SPACEV1B}
\addplot[dataset gmm] table[
    x expr=\thisrow{total_inserted}/1000000,
    y expr=100*\thisrow{ghost_memory_mb}/\thisrow{rss_before_mb},
] {\ghostgmm};
\addlegendentry{\gmm{1B}}
\end{axis}
\node[mem subcap] (d-cap) at (d.south |- d.outer south)
    {\subcaption{Ghost edges}\label{fig:eval:memory:ghost}};
\end{tikzpicture}

\vspace{-2mm}
\caption{Memory consumption of \system\ as the indexed collection grows.
(a)--(c): peak memory per worker (line: mean; band: min--max across workers);
the dashed line marks the memory of one node.
(d): ghost-edge memory relative to index memory, summed over all workers except the first and last phases.
As the index grows 10$\times$ to 1B vectors, the mean per-worker peak memory grows only 1.8$\times$ and stays within 83\% of a node's RAM. Ghost edges add 11--16\% memory overhead from 400M vectors onward.
}
\label{fig:eval:memory}
\end{figure}


This experiment aims to show that \system keeps per-worker memory bounded as the index grows, with a low memory overhead from ghost edges.

\paragraph{Setup.}
We apply the same incremental index building setup as \cref{sec:eval:build}. For each worker, we track the peak memory usage reached since its startup; this value is read at the end of repartitioning in each phase. We also measure the ghost edges memory overhead, by recording the memory consumption right before and after loading ghost edges for each worker. So the ghost edges overhead is the total extra memory that loading adds on all workers, divided by their total memory before loading.

\paragraph{Results.}
\cref{fig:eval:memory:sift}--\cref{fig:eval:memory:gmm} show the trends of per-worker peak memory as the index grows on three datasets. The highest per-worker peak across all phases is 207\,GiB on SIFT1B, 213\,GiB on SPACEV1B and 204\,GiB on \gmm{1B}, at most 83\% of the 256\,GiB node RAM; this peak is reached in the repartitioning after 800M vectors (600M on \gmm{1B}). From 100M to 1B vectors, the mean per-worker peak grows about 1.8$\times$ from 84--90\,GiB to 153--166\,GiB, while the dataset grows by 10$\times$.
\cref{fig:eval:memory:ghost} shows the memory overhead of ghost edges. From 400M vectors onward, ghost edges add 11--16\% to the workers' memory, and this does not grow as workers join.

\paragraph{Analysis.}
A worker usually reaches its peak memory while merging the index pieces it receives, at a median of 50--60\,GiB above its memory at the end of insertion.
As described in \cref{sec:scale-liveness}, while merging, the worker still keeps its old index to serve queries, in addition to the staged index and the index piece it is merging.
This extra memory is released once the old index is retired.
The share of ghost edges in memory stays about the same: although ghost edges keep growing (\cref{sec:eval:rebalancing_efficiency}), the index grows at the same pace.

\section{Related Work}
\label{sec:related_work}

We organize related work into single-node vector indexes whose graph structures, streaming update mechanisms, and clustering designs inform our local worker architecture (\cref{sec:related_single}), and distributed vector search systems spanning sharded, global-graph, and hybrid partitioning paradigms (\cref{sec:related_dist}).

\subsection{Single-Node Systems}
\label{sec:related_single}

\paragraph{Graph-Based Indexes.}
Graph-based indexes answer queries by greedy traversal over neighbor links and lead in-memory ANNS in recall and throughput~\cite{aumuller2020annbenchmarks, wang2021comprehensivesurveyexperimentalcomparison, graphvector}.
In-memory designs include NSG~\cite{nsg} and HNSW~\cite{hnsw}, while disk-resident designs such as DiskANN~\cite{diskann} and Starling~\cite{wang2024starling} store the graph on SSD to reach billion scale on one machine.
These systems provide the foundational index structure that \system distributes across workers.


\paragraph{Incremental Indexing.}
Streaming graph indexes either buffer insertions and periodically merge them into the on-disk graph~\cite{freshdiskann} or update the graph in place~\cite{odinann, xu2025inplaceupdatesgraphindex}.
Deletions and updates further require repairing neighbor links, since frequent updates otherwise leave vertices unreachable~\cite{xiao2024enhancinghnswindexrealtime, xu2025inplaceupdatesgraphindex}.
These repairs are confined to a single graph on one machine. \system's ghost edges instead preserve cross partition links as vectors migrate between workers.


\paragraph{Cluster-Based Indexes.}
IVF-style indexes such as SPANN~\cite{spann} keep centroids in memory and posting lists on disk.
Because vectors in different clusters share no edges, these indexes can cheaply split and merge clusters online to handle updates and query skew~\cite{spfresh, adaivf, quake}.
\system adopts a similar split/merge pattern, inspired by SPFresh's LIRE~\cite{spfresh}, but must additionally preserve graph edges that cross partition boundaries.


\subsection{Distributed Systems}
\label{sec:related_dist}

\paragraph{Sharded Approaches.}
Sharded approaches partition the dataset and build an independent index per partition~\cite{lanns, pyramid, elpis}.
Their recall depends on partitioning and routing, which prior work improves by co-locating near neighbors~\cite{dong2020learningspacepartitionsnearest, query_router_graph_partitioning, bliss} and by routing with centroid indexes~\cite{pyramid} or error bounds~\cite{Auncel}.
However, they assume static partitions, whereas \system repartitions dynamically and keeps its router up-to-date via reservoir sampling~\cite{reservoir_sampling}.


\paragraph{Global Graph Approaches.}
Global graph approaches build one graph over all vectors and spread it across machines~\cite{distributedann, cotra, yu2025approximatenearestneighborsearch} or a disaggregated memory pool~\cite{cxl-anns}.
Search hops therefore cross machines. Prior work reduces the communication cost with RDMA~\cite{cotra}, asynchronous I/O~\cite{yu2025approximatenearestneighborsearch}, or near-data processing~\cite{cxl-anns}.
However, remote hops remain on the search path. \system instead keeps each local search within one worker and retains cross-worker edges as ghost edges without traversing them.


\paragraph{Hybrid Approaches.}
SPIRE~\cite{spire} introduces a hierarchical IVF structure with accuracy-preserving recursive construction. It identifies a balanced partition granularity to avoid read-cost explosion. However, SPIRE uses a proximity graph only at the root level to route queries, while its leaf partitions are IVF-style posting lists on SSD that are scanned rather than traversed.
Harmony~\cite{harmony} takes a different approach, combining dimension-based and vector-based partitioning to balance workloads under skewed query distributions. Its cost model adaptively selects partitioning strategies per query, achieving high throughput across diverse workloads. However, Harmony uses static partitioning without support for dynamic repartitioning or incremental updates.


\section{Conclusion}
\label{sec:conclude}

We presented \system, a distributed graph-based ANNS system that achieves horizontal elasticity without full index reconstruction. By introducing \emph{ghost edges} to retain cross-partition proximity relationships on disk with zero search-time overhead, \system enables an online \emph{split-and-merge} repartitioning protocol that reactivates dormant edges as partition boundaries shift, paired with a lightweight reservoir-sampled seed index for selective query routing. Across three billion-scale datasets, \system scales incrementally from 1 to 10 workers with stable per-phase insertion times while existing systems time out or exhaust memory, delivers up to 3.9$\times$ and 27$\times$ higher query throughput than \textsf{SPTAG} and \textsf{Harmony} at Recall@10 of 0.99, achieves 2.4--3.0$\times$ throughput scale-up with consistent accuracy, and sustains query throughput across repartitioning with at most 16\% ghost-edge memory overhead.



\bibliographystyle{ACM-Reference-Format}
\bibliography{_refs}

\end{document}